\pdfoutput=0

\documentclass[12pt,onecolumn,journal]{IEEEtran}
\usepackage{amsmath,amssymb,amsthm}
\usepackage[noadjust]{cite}
\usepackage{gensymb}
\let\labelindent\relax
\usepackage[font=footnotesize]{caption}
\usepackage{subcaption}
\usepackage{xspace,arydshln}
\usepackage{graphicx,fancyhdr,epstopdf}
\graphicspath{{./figures/}}
\usepackage{enumitem}
\usepackage[table]{xcolor}
\usepackage{multirow}
\usepackage{comment}
\usepackage{steinmetz} % For the phase symbol with the \phase{} command
\usepackage{gensymb} %For \degree
\usepackage{mathrsfs}%${\mathscr F} {\mathscr L}$ %Real Math Script
\usepackage{siunitx}
\usepackage{algorithm}
\usepackage{algpseudocode}
\begin{document}
\begin{comment}
%
\title{AI-Driven Optimization of Wave-Controlled Reconfigurable Intelligent Surfaces}
%
\author{GAL BEN-ITZHAK (Graduate Student Member,~IEEE),
MIGUEL SAAVEDRA-MELO (Graduate Student Member,~IEEE),
ENDER AYANOGLU (Fellow,~IEEE),
FILIPPO CAPOLINO (Fellow,~IEEE), AND
A. LEE SWINDLEHURST (Fellow,~IEEE)
}
       % <-this % stops a space
%\affil{Center for Pervasive Communications and Computing, Department of Electrical Engineering and Computer Science, University of California, Irvine, Irvine, CA 92697 USA}
%\corresp{CORRESPONDING AUTHOR: Ender Ayanoglu (e-mail: ayanoglu@uci.edu).}
%\authornote{This work is partially supported by NSF grant ECCS-2030029.}
%\markboth{AI-Driven Operation of Wave-Controlled Reconfigurable Intelligent Surfaces}{Ayanoglu \textit{et al.}}
\end{comment}
\title{AI-Driven Optimization of Wave-Controlled Reconfigurable Intelligent Surfaces
%{\footnotesize \textsuperscript{*}Note: Sub-titles are not captured in Xplore and
%should not be used}
\thanks{The authors are with the Center for Pervasive Communications and Computing (CPCC), Department of Electrical Engineering and Computer Science,
University of California, Irvine.}
\thanks{This work is partially supported by NSF grant 2030029.}
}

\author{\IEEEauthorblockN{Gal Ben Itzhak, {\em Graduate Student Member, IEEE,}}\\
\IEEEauthorblockN{Miguel Saavedra-Melo, {\em Graduate Student Member, IEEE,}}\\
\IEEEauthorblockN{Ender Ayanoglu, {\em Fellow, IEEE,}}\\
\IEEEauthorblockN{Filippo Capolino, {\em Fellow, IEEE,}}\\
\IEEEauthorblockN{and A. Lee Swindlehurst, {\em Fellow, IEEE}}\\
}

\maketitle
\begin{abstract}
A promising type of Reconfigurable Intelligent Surface (RIS) employs tunable control of its varactors using biasing transmission lines below the RIS reflecting elements. Biasing standing waves (BSWs) are excited by a time-periodic signal and sampled at each RIS element to create a desired biasing voltage and control the reflection coefficients of the elements. A simple rectifier can be used to sample the voltages and capture the peaks of the BSWs over time. Like other types of RIS, attempting to model and accurately configure a wave-controlled RIS is extremely challenging due to factors such as device non-linearities, frequency dependence, element coupling, etc., and thus significant differences will arise between the actual and assumed performance.
%Configuring the RIS to provide a certain radiation pattern is a complicated task. The entire RIS system is complex and nonlinear, from the electrical behavior of the BSWs at different frequencies, to the coupling between RIS elements, and to the channel models of the environment, and is therefore difficult to model the entire system with simulations and optimize the resulting radiation patterns for given RIS configurations. 
An alternative approach to solving this problem is data-driven: Using training data obtained by sampling the reflected radiation pattern of the RIS for a set of BSWs, 
%As the RIS is deployed, it is configured by different random sets of BSWs on the biasing transmission line, with the corresponding received power measured in the directions of interest for each set of BSW amplitudes. 
a neural network (NN) is designed to create an input-output map between the BSW amplitudes and the resulting sampled radiation pattern. This is the approach discussed in this paper. In the proposed approach, the NN is optimized using a genetic algorithm (GA) to minimize the error between the predicted and measured radiation patterns. The BSW amplitudes are then designed via Simulated Annealing (SA) to optimize a signal-to-leakage-plus-noise ratio measure by iteratively forward-propagating the BSW amplitudes through the NN and using its output as feedback to determine convergence. The resulting optimal solutions are stored in a lookup table to be used both as settings to instantly configure the RIS and as a basis for determining more complex radiation patterns.
\end{abstract}
\maketitle
\section{INTRODUCTION}
A Reconfigurable Intelligent Surface (RIS) is a promising building block for next-generation wireless networks, potentially solving issues related to limited coverage, sensitivity to blockages, high path loss, etc., by providing alternative propagation paths and enabling beamforming for desired directions \cite{9475160}. An RIS is a type of metasurface \cite{6331512, 9887248} consisting of periodic arrangements of subwavelength-sized passive reflecting elements whose reflection coefficients can be individually controlled. 

\begin{figure}
    \centering
    \includegraphics[width=1.0\linewidth]{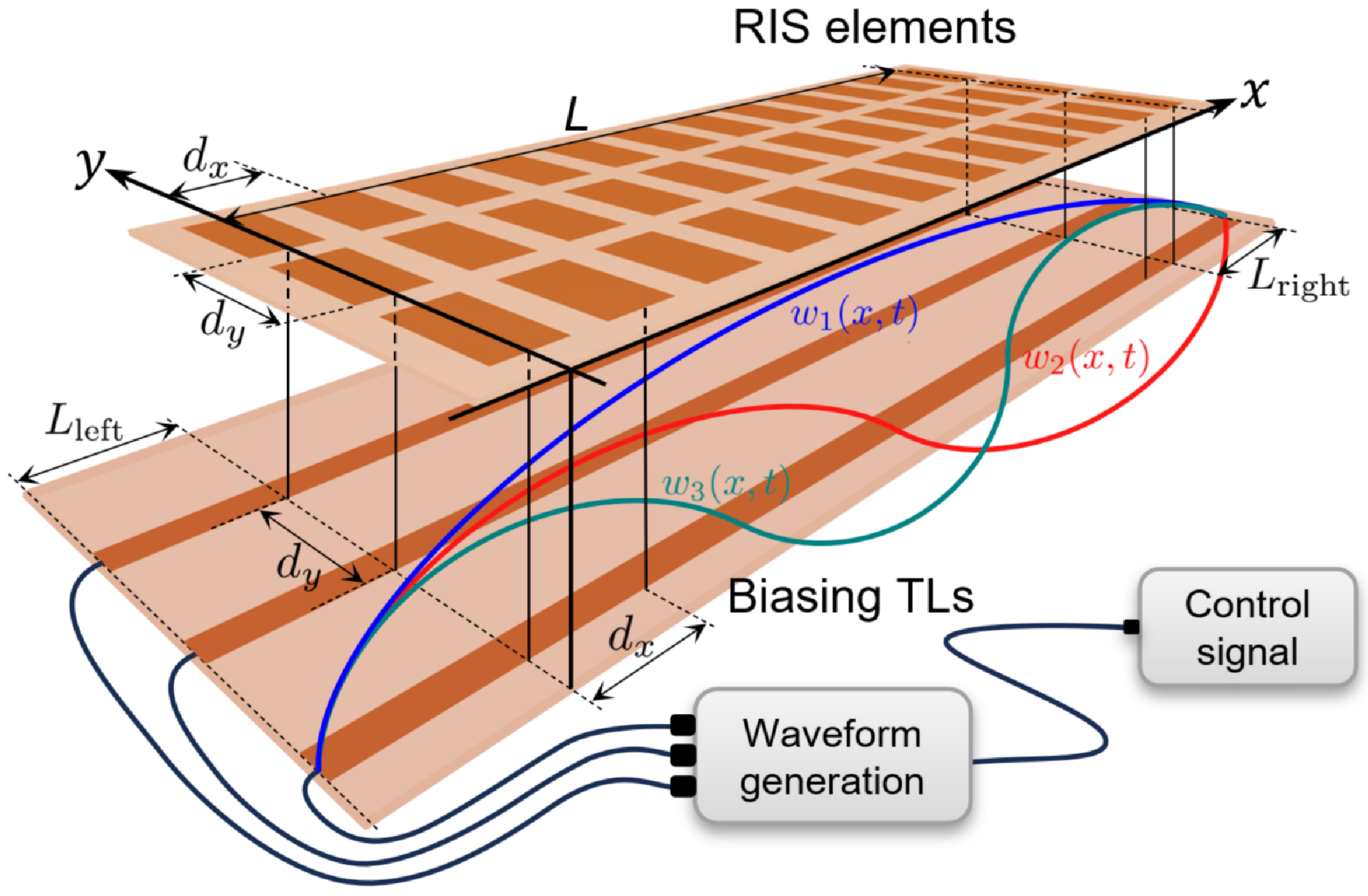}
    \caption{Wave-controlled RIS composed of two physical layers \cite{10742896}. Top layer: $M$ elements in each row along the $x$ direction, where each element is connected to a varactor diode. Bottom layer: $N$ BSWs are excited on each TL to create the biasing voltages sampled at each RIS element. Each row is controlled via the connection on the left where a signal with $N$ adjustable frequency components is injected by a waveform generator. Adjacent metallic patches and varactors on the top layer, and adjacent DC voltage outputs on the bottom layer are uniformly separated by distance $d_x$ in the $x$ direction and $d_y$ in the $y$ direction.}
    \label{fig:wave_controlled}
\end{figure}

The RIS architecture studied in this paper is illustrated in Fig.~\ref{fig:wave_controlled}, and is based on the general wave-controlled approach described in \cite{9770088}. In this approach, biasing standing waves (BSWs) are excited on a transmission line (TL) behind the RIS structure and sampled at discrete locations to provide the DC bias voltages for varactor diodes that modulate the reflection coefficients. In a previous paper \cite{10742896}, various optimization algorithms were developed that demonstrate the capability of the wave-controlled RIS approach to create beams and nulls in the radiation pattern for multiple desired directions. The performance of the optimization algorithms were demonstrated for two different types of voltage sampling circuits: a {\em rectifier} that is easier to implement but more difficult to use for optimization, and a {\em sample-and-hold} circuit, which is more costly to implement but enables analytical solutions to the  beam pattern optimization problem. 

In this work, we consider the rectifier implementation, where the beam pattern optimization is nonlinear and constrained, and we assume the RIS is deployed in an unknown environment without knowledge of channel state information (CSI). Thus, there is no exact mathematical model that can be used to predict the response created by the RIS to an excitation by the BSWs. Consequently, we will turn to the use of machine learning (ML) tools to solve the RIS configuration problem.

%The separation between adjacent metallic patches is $\Delta_x$ in the $x$ direction and $\Delta_y$ in the $y$ direction.

There exist many algorithms for RIS optimization in the literature, such as model-based, ML-based, heuristic, and combinations of ML / model-based and heuristic algorithms \cite{10361836}. Model-based optimization in the physical environment is difficult, in particular because of arbitrary control of the individual RIS reflection coefficients not being possible due to the 
coupling between the RIS elements. Moreover, typically available mathematical models for the RIS behavior do not take into account factors such as frequency dependence or device nonlinearities. Many of the supervised ML algorithms in \cite{10361836} have been demonstrated to work well for energy efficiency and data rate maximization tasks, though they often make assumptions about existing knowledge of the channels, which is limited in practice given the complexity of CSI acquisition with the passive RIS. There is also difficulty in architecting the neural networks for system modeling while avoiding overfitting, with some solutions including adding dropout layers (randomly ignoring neuron contributions) \cite{9380744}, using larger datasets, and decreasing the number of hidden layers \cite{9448826}. Though, the model performances are very sensitive to hyperparameter selection and will vary on a case-to-case basis, requiring time and experience in fine-tuning. It is desirable to have a systematic way to generate and evaluate the performance of models without manually optimizing them. In \cite{9317827}, a deep-learning convolutional neural network was demonstrated to capture the relationships between RIS configurations and achievable rates at arbitrary receiver (Rx) locations, and an optimal set of discrete RIS phase shifts was recommended to optimize the propagation environment. This is done by an exhaustive search on an existing codebook of RIS phase shifts during inference, which adds additional delays. Different learning algorithms, such as unsupervised learning, are also feasible, as shown for the case of signal-to-noise ratio (SNR) maximization in the setting of a single user \cite{8955968}. However, the size of the training dataset required for optimal network convergence increases significantly with the number of RIS elements, posing a challenge for the scalability of the RIS.

%The use of ML for RIS-based passive beamforming has been widely explored in the literature. \cite{10716384, 10361836, 8955968, 9451567, 9317827, 9013256}. In \cite{10612784} and \cite{10694808}, blind beamforming for specific directions was demonstrated based on real-time and iterative feedback from the receivers for learning and rapid optimization, but such feedback may not be available. Most papers, however, make assumptions on the system that may not hold in practical scenarios. In particular, arbitrary control of the individual RIS reflection coefficients is not possible due to coupling between RIS elements, knowledge of the transmitter-RIS and RIS-receiver channels is limited due to the complexity of CSI acquisition with the passive RIS, and typically available mathematical models for the RIS behavior do not take into account factors such as frequency dependence or device nonlinearities.  

In the approach presented in this paper, we address these shortcomings. In particular, we focus on the use of wave-controlled RIS, which as described above is a novel architecture that accommodates the physical limitations of an RIS and simplifies the hardware required to operate it.  The wave-controlled RIS achieves reduced-dimension control of the $M$ RIS elements by configuring $N \ll M$ BSW amplitudes on a TL. This approach also accounts for the coupling between RIS elements by maintaining a relatively smooth variation in the voltage profiles across adjacent elements. Further, we develop an ML model to not only approximate the behavior of the wave-controlled RIS in a given environment, but also to optimize its radiation pattern. Using the data-driven approach discussed in this paper, CSI for the transmitter-RIS and RIS-receiver links is not required for configuration of the RIS. In particular, we propose to optimize the RIS through supervised learning and offline optimization via three different algorithms: backpropagation, a genetic algorithm (GA), and simulated annealing (SA). The RIS is initially excited by random sets of BSWs, and the corresponding radiation patterns are sampled at specific directions of interest. Then, a neural network (NN) is proposed to model the relationship between the amplitudes of the BSWs and the corresponding power samples, and the network architecture is optimized using the GA to create a sufficiently generalized mapping between the two. SA is then used to optimize the amplitudes and produce the desired radiation pattern, using the received power estimates from the NN as feedback during the iterations of the optimization algorithm. Finally, optimal BSW configurations are stored in a lookup table that can be accessed for both rapid RIS beamsteering and as bases for optimizations involving more complex radiation patterns. 

The paper is organized as follows. In Section II, we discuss the electromagnetic and circuit models used to simulate the wave-controlled RIS in a physical environment. Section III discusses the purpose of using an ML model to control the RIS and outlines procedures for collecting data from the environment to optimize the ML model using a GA. In Section IV, we show the interoperability of SA and the ML model to optimize beam patterns for various objective functions. Finally, Section V analyzes the performance gain achieved by employing a lookup table to store data used to configure the RIS for different radiation patterns.

% This paper proposes the novel approach of RIS optimization through supervised learning with measurements collected from receivers in directions of interest, as the RIS is excited with random sets of standing waves. The architecture of the neural network used for the envelope-detector RIS and trained with back-propagation, is optimized using a genetic algorithm to achieve a good general representation of the RIS and channel models. The neural network is then used with simulated annealing and an evolving lookup table to optimize radiation patterns to generate multiple beams and nulls in arbitrary directions. As the lookup table grows, the RIS will require fewer offline optimizations and can refer to existing configurations for rapid beam-steering.
\section{MODEL, ASSUMPTIONS, AND NOTATION}
\subsection{RIS MODEL}
\begin{figure}
    \centering
    \includegraphics[width=0.8\linewidth]{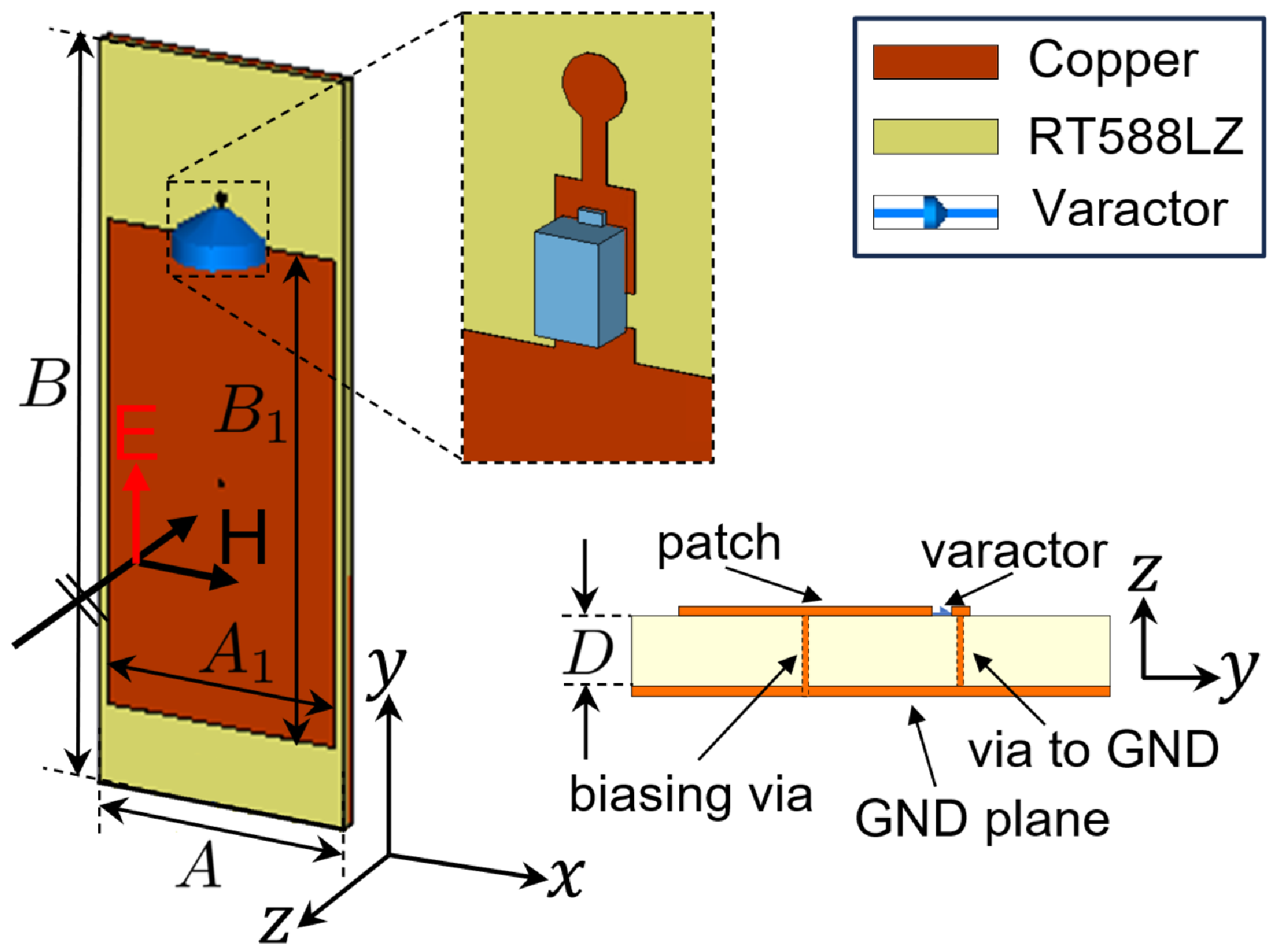}
    \caption{RIS unit cell geometry. Each rectangular metallic conductor is biased by the sampled voltage through a via and connected to a grounded varactor. The reverse-biased varactors act as tunable capacitors to polarize the incident electric field along the $y$ direction.}
    \label{fig:grounded_varactor}
\end{figure}

In Fig.~\ref{fig:wave_controlled}, we show an RIS with uniform linear arrays of $M$ metallic patches in the $x$ direction with period $d_x$. The figure shows three parallel linear arrays as an example. Each RIS element is connected to a varactor diode which is used to vary the reflective properties of the element, as illustrated in Fig.~\ref{fig:grounded_varactor}. On the backside of the structure, there is a biasing TL of length $L$ along $x$ on which a set of $N$ BSWs are excited over the entire length of the TL. We assume that the fundamental standing wave is resonant with the total TL length $L_{\text{tot}}$, leading to the fundamental angular frequency $\omega_{\text{b}} = 2\pi f_{\text{b}}=\pi v_{\text{ph}}/L_{\text{tot}}$ \cite{10237459}. The other BSWs oscillate at the harmonic frequencies $n f_{\text{b}}$, with $n=1,2,\ldots,N$. The overall BSW signal, assuming a short circuit termination at the end of the biasing TL, is given by
\begin{equation}
    w(x,t) = W_0 + \sum_{n=1}^{N} W_n  \sin\left(\frac{n \pi (x+L_{\text{left}})}{L_{\text{tot}}}\right) \sin (n \omega_{\text{b}} t ),
    \label{eqn:Wxt}
\end{equation}
where the individual BSWs are controlled by tuning their amplitudes, which are stored in the vector $\boldsymbol{W}=[W_1, W_2, \ldots, W_N]$. The $W_0$ term is a DC component used to center the varactors' biasing on their best working range for capacitance control.

\begin{figure}
    \centering
    \includegraphics[width=0.8\linewidth]{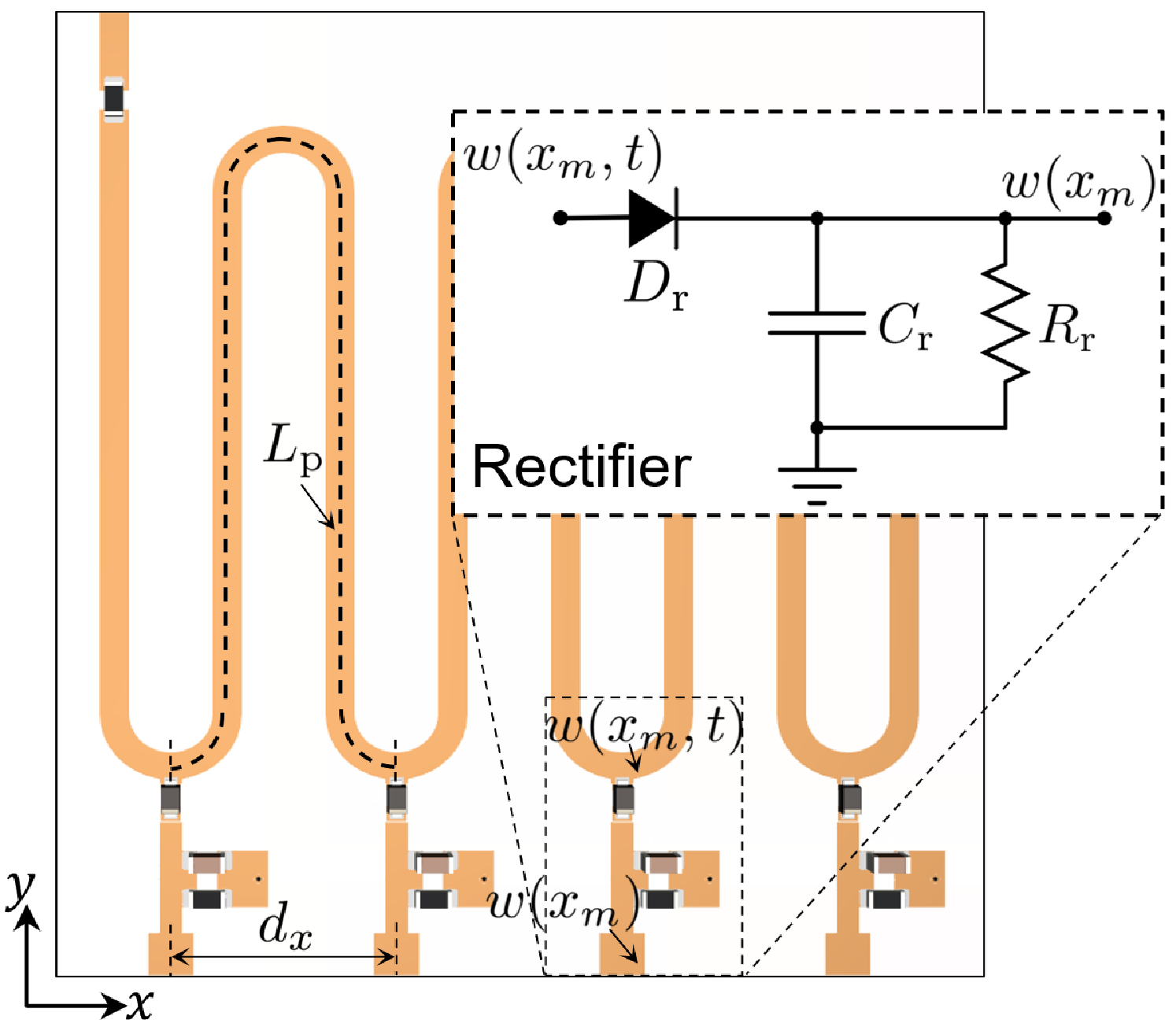}
    \caption{Geometry of the biasing TL. The length $L_\mathrm{p}$ of the TL path for one unit cell and the distance $d_x$ between two adjacent rectifier circuits are detailed. These circuits, which rectify the BSWs \cite{10742896}, are located at the bottom, with one assigned to each RIS element. The voltage $w(x_m,t)$ is extracted from the biasing TL at the location $m$, while the rectified voltage $w(x_m)$ provides the bias to the varactor of the $m$th element. $w(x_m,t)$ is rectified using the diode $D_\mathrm{r}$ and by following its envelope or peaks through the $RC$ circuit shown, with a carefully chosen time constant to minimize voltage drops in $w(x_m)$ due to capacitor discharge.}
    \label{fig:TL_geometry_Rectifier}
\end{figure}

To minimize the control signaling overhead and reduce the variation of the bias voltage from element to element, it is desired that $N \ll M$. Connected to each RIS element with a vertical via is a rectifier circuit, shown in Fig.~\ref{fig:TL_geometry_Rectifier}, which rectifies the BSW signal at each RIS element location $x_m=md_x$, with $m=0,1,2,..M-1$. We define the reference point $x=0$ at the location where the first rectifier is located and $L_{\text{left}}$ as the length of the biasing TL in the $x$ direction before (to the left of) this detector. We also define $L_{\text{right}}$ as the length of the biasing TL between the last rectifier and the short circuit at the end of the TL on the right. The total length of the waveguide along $x$ is given by
\begin{equation}
    L_{\text{tot}}=L_{\text{left}}+L+L_{\text{right}}.
\end{equation}

The voltage variation along $x$ for each BSW is based on the phase velocity  $v_{\text{ph}}$, which we assume due to low frequency dispersion to be the same for all the waves. The phase velocity is given by $v_{\text{ph}}=c/n_{\text{slow}}$, where $c$ is the speed of light and $n_{\text{slow}}$ is the slowness factor that depends on the geometry and material properties of the biasing TL. As shown in Fig.~\ref{fig:TL_geometry_Rectifier}, $L_{\text{p}}$ represents the physical path length between two adjacent DC voltage outputs along the biasing TL, while $d_x$ is the corresponding separation along the $x$ direction. The effective refractive index, $n_{\text{eff}}$, of the biasing TL is determined by its effective permittivity $\epsilon_{\text{eff}}$ as $n_{\text{eff}} = \sqrt{\epsilon_{\text{eff}}}$. Consequently, the slowness factor in this case is given by $n_{\text{slow}}=(L_{\text{p}}/d_x)n_{\text{eff}}$. 

The voltage is uniformly sampled at discrete locations $x_m = m d_x$, where $m=0,1,\ldots,M-1$, and $0 \leq x_m \leq L$.
We rewrite~(\ref{eqn:Wxt}) in terms of the RIS element indices $m$ as
\begin{equation}
\begin{split}
    & w (md_x,t) = \\
    & W_0 + \sum_{n=1}^{N}{W_n \sin\left(\frac{n \pi (m d_x+L_{\text{left}})}{L_{\text{tot}}}\right)\sin(n\omega_{\text{b}} t)}.
%   \label{eq:sample_and_hold_wave_model}
\end{split}
    \label{eqn:wmt1}
\end{equation}
The varactor diode connected to each RIS element is biased in reverse using the rectifier circuit as in the inset of Fig.~\ref{fig:TL_geometry_Rectifier}.
We assume that the rectified voltage of the BSWs at each discrete location $x_m$ is equivalent to the peak voltage (maximum voltage at each location over time), yielding
\begin{equation}
\begin{split}
    & w(md_x) = \\
    & W_0+\max_t{\left(\sum_{n=1}^{N}{W_n \sin\left(\frac{n \pi (md_x + L_{\text{left}})}{L_{\text{tot}}}\right)\sin(n\omega_{\text{b}} t)}\right)},
    \label{eq:envelope_detector_wave_model}
\end{split}
\end{equation}
\begin{comment}
\begin{equation}
\begin{split}
    & w(mL_\text{p}) = \\
    & W_0+\max_t{\left(\sum_{n=1}^{N}{W_n \sin\left(\frac{n \pi (mL_\text{p} + L_{\text{left}})}{L+{\text{tot}}}\right)\sin(n\omega_{\text{b}} t)}\right)},
    \label{eq:envelope_detector_wave_model}
\end{split}
\end{equation}
\end{comment}
\begin{comment}
\begin{equation}
\begin{split}
    & w(m) = \\
    & W_0+\max_t{\left(\sum_{n=1}^{N}{W_n \sin\left(\frac{n \pi (m+M_l)}{M-1+M_l+M_r}\right)\sin(n\omega_{\text{b}} t)}\right)},
    \label{eq:envelope_detector_wave_model}
\end{split}
\end{equation}
\end{comment}
with $W_0$ taken out of the $\max_t(\cdot)$ term as it is a constant independent of time.
This positive voltage is then applied at the cathode of the varactor, thus reverse-biasing it, determining its capacitive properties, and consequently tuning the reflection coefficient of the RIS element to provide the desired response.
\begin{comment}
\begin{equation}
    w (mL_\text{p},t) = 
    W_0 + \sum_{n=1}^{N}{W_n \sin\left(\frac{n \pi (m L_\text{p}+L_{\text{left}})}{L_{\text{tot}}}\right)\sin(n\omega_{\text{b}} t)}.
%   \label{eq:sample_and_hold_wave_model}
    \label{eqn:wmt1}
\end{equation}
\begin{equation}
\begin{split}
    w & (md_x,t) = \\
    & W_0 + \sum_{n=1}^{N}{W_n \sin\left(\frac{n \pi (m+M_l)}{M-1+M_l+M_r}\right)\sin(n\omega_{\text{b}} t)},
%   \label{eq:sample_and_hold_wave_model}
    \label{eqn:wmt1}
    \end{split}
\end{equation}
where $M_l=L_{\text{left}}/d_x$ and $M_r=L_{\text{right}}/d_x$ are unitless quantities.
\end{comment}
%\begin{figure}
%    \centering
%    \includegraphics[width=0.75\linewidth]{figures/envelopedetector.eps}
%    \caption{Envelope detector circuit used to rectify the alternating current voltage on the biasing TL \cite{10742896}. The input voltage $v_i(t)$ is the standing waves at location $m$; the output voltage $v_o(t)$ is the rectified voltage to reverse-bias the varactor at location $m$. The circuit rectifies $v_i(t)$ using the diode $D$ and follows its envelope or peak through the $RC$ circuit shown, with a carefully chosen $RC$ time constant that allows to follow the peak without significant drops in $v_o(t)$ due to capacitor discharge.}
%    \label{fig:envelope_detector}
%\end{figure}

\begin{figure}
    \centering
    \includegraphics[width=0.7\linewidth]{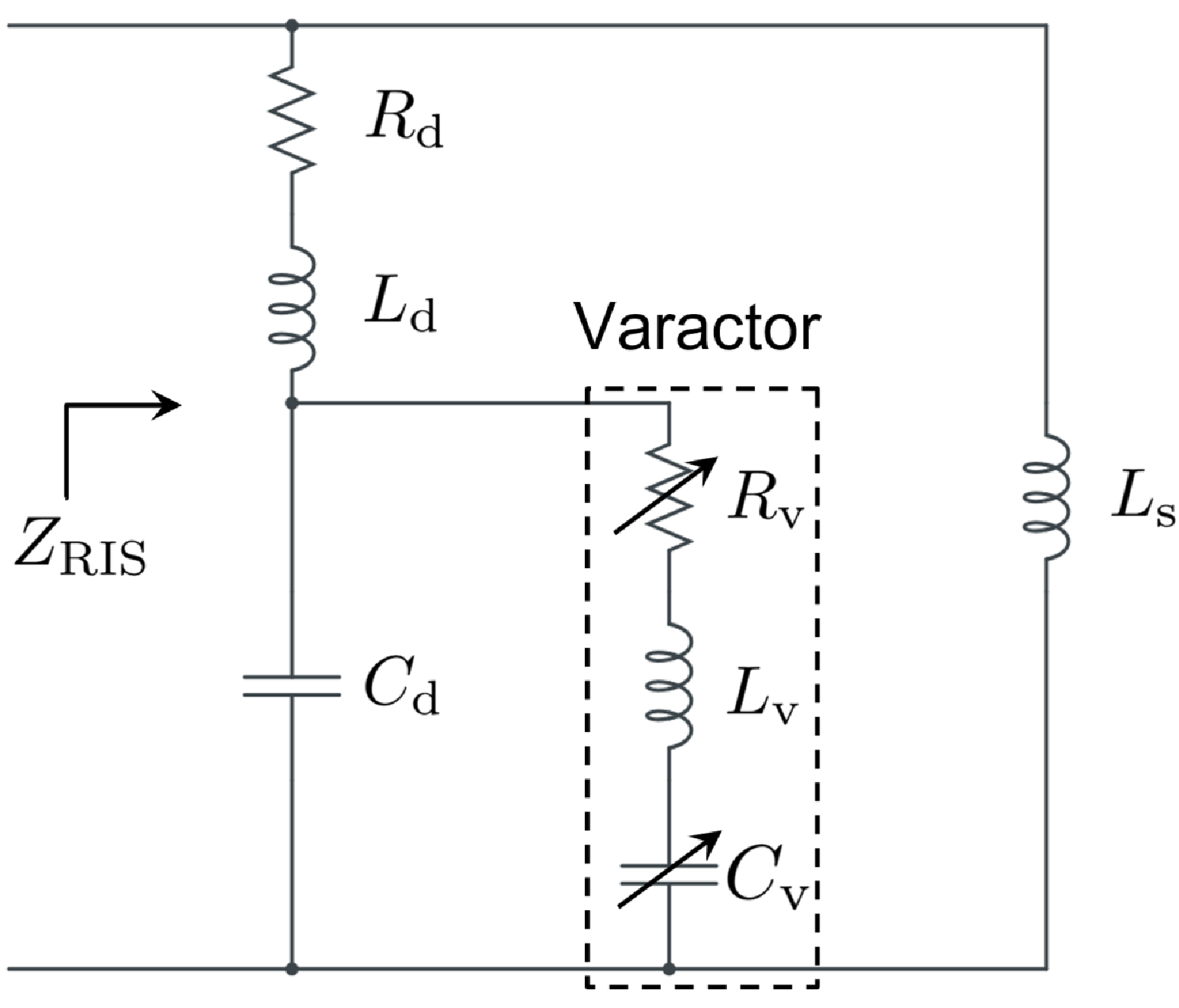}
    \caption{Analytical model of the RIS unit cell including an RLC model of the varactor to calculate the reflection coefficient as a function of frequency or varactor biasing voltage. The equivalent impedance $Z_{\text{RIS}}$ is used to find the reflection coefficient $\Gamma$.}
    \label{fig:circuit_ris}
\end{figure}

Each RIS unit cell shown in Fig.~\ref{fig:grounded_varactor} is designed with (all dimensions in mm) $A=20,\: A_1=18.5,\: B=72.7,\: B_1=36.7,\: D=1.27$. The dielectric spacer is the substrate Rogers RT5880LZ, with relative permittivity $\epsilon_r=2$ and dielectric loss $\tan(\delta)=0.0021$. To simulate the tunable impedance of each RIS element, the equivalent circuit model shown in Fig.~\ref{fig:circuit_ris} is used and optimized with full-wave simulations \cite{Costa21, 9887248}.  The unit cell elements have resistance $R_{\text{d}}=0.1671$ $\Omega$, capacitance $C_{\text{d}}=0.97821$ pF, and inductance $L_{\text{d}}=1.9177$ nH. There is an additional inductive term $L_{\text{s}}=1.5959$ nH in parallel to the unit cell element and varactor that accounts for the grounded substrate, creating the ``magnetic resonance" effect discussed in \cite{6331512, Sievenpiper99, Best08}.

\begin{figure}
    \centering
    \includegraphics[width=0.9\linewidth]{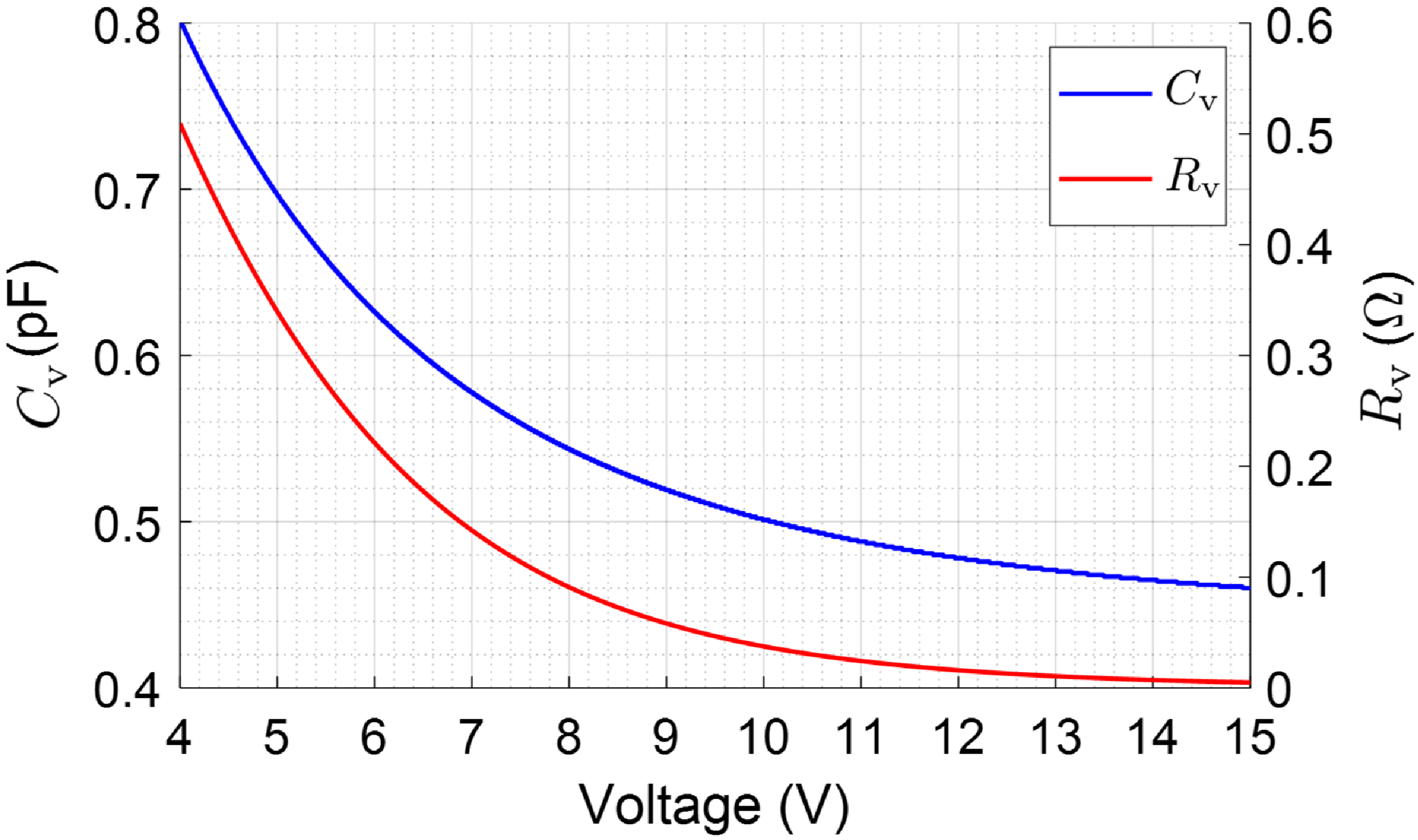}
    \caption{Equivalent resistance and capacitance for the SMV1231-040LF varactor as functions of the biasing voltage across the varactor model shown in Fig.~\ref{fig:circuit_ris}.}
    \label{fig:varactor_c_r}
\end{figure}

The varactor SMV1231-040LF provided by Skyworks Solutions, Inc. is chosen due to its desirable characteristics, including resistance below $0.6$ $\Omega$ and low series inductance ($0.45$ nH) for our frequencies of interest. The varactor is modeled by an equivalent series RLC circuit with inductance $L_{\text{v}}$, capacitance $C_{\text{v}}\text{(V)}$, and resistance $R_{\text{v}}\text{(V)}$. The inductance $L_{\text{v}}=2.34$ nH is constant and models the package and parasitic inductance when the varactor is connected across the gap. The values for $C_{\text{v}}\text{(V)}$ and $R_{\text{v}}\text{(V)}$ shown in Fig.~\ref{fig:varactor_c_r} are obtained from a parametric sweep simulation using Advanced Design System (ADS) software, as functions of the reverse voltage bias across the varactor model provided by the vendor, limited to the range $[4\ \text{V}, 15\ \text{V}]$. 

\begin{figure}
    \centering
    \includegraphics[width=0.9\linewidth]{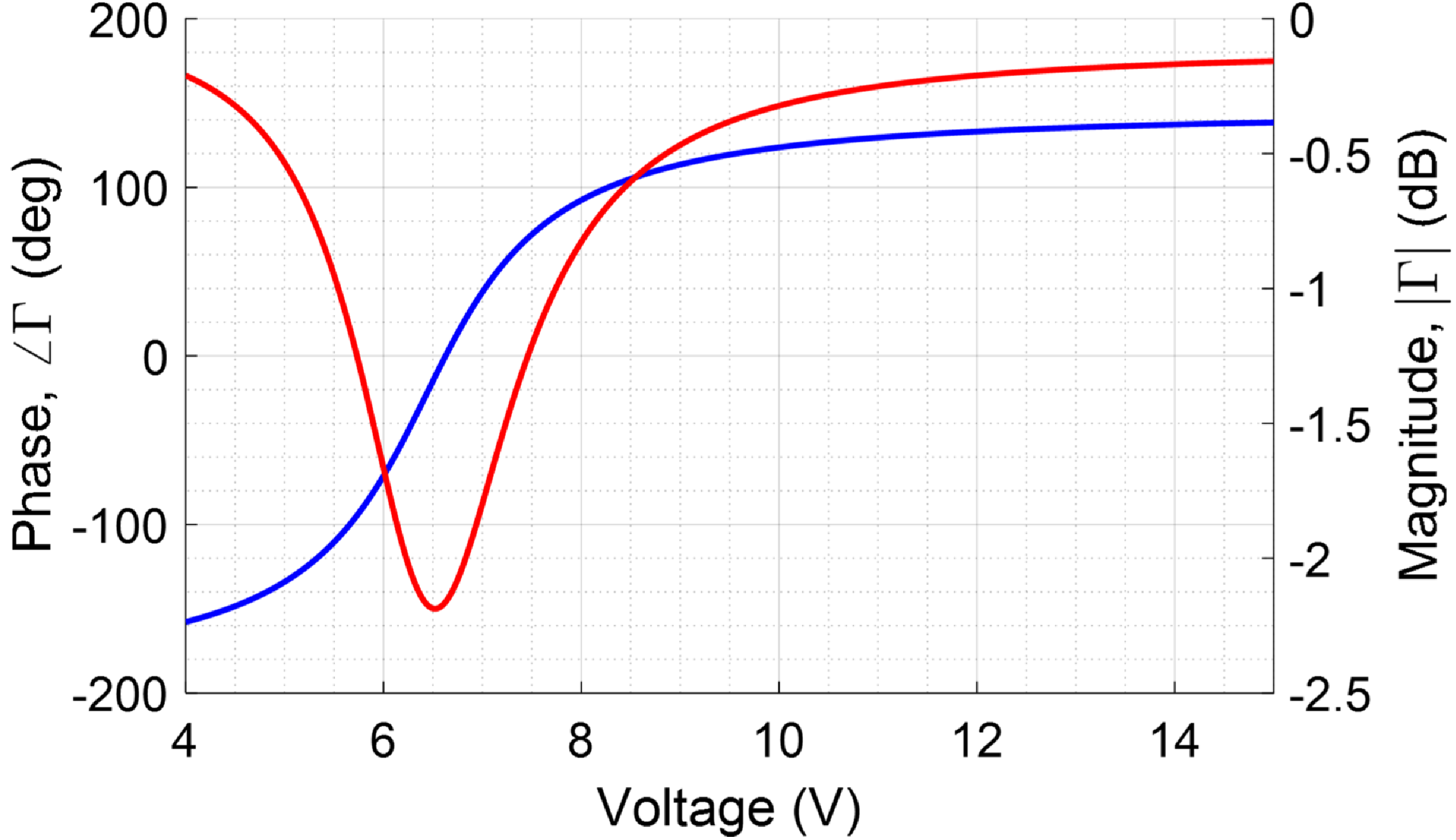}
    \caption{Reflection magnitude and phase of the RIS unit cell described in Fig.~\ref{fig:grounded_varactor} as functions of the reverse bias voltage across the varactor at the carrier frequency $f_{\text{c}}=2.45$ GHz.}
    \label{fig:reflection_coeff}
\end{figure}

The total equivalent RIS impedance is expressed by
\begin{equation}
\begin{split}
    & Z_{\text{RIS}}= \\
    & \left(R_{\text{d}}+j\omega L_{\text{d}} + \left(R_{\text{v}} + j\omega L_{\text{v}} + \frac{1}{j\omega C_{\text{v}}}\right)\parallel\frac{1}{j\omega C_{\text{d}}}\right)\parallel j\omega L_{\text{s}}, 
\end{split}
\end{equation}
and is used to determine the reflection coefficient of the RIS elements by
\begin{equation}
    \Gamma = \frac{Z_{\text{RIS}}-Z_0}{Z_{\text{RIS}}+Z_0},
\end{equation}
where $Z_0$ is the free-space impedance. This model predicts the magnitude and phase of the local reflection coefficient for variable frequency and varactor biasing voltage $V$. An example is given in   Fig.~\ref{fig:reflection_coeff}, where the magnitude and phase of the reflection coefficients as functions of the varactor reverse bias voltage are plotted for the frequency $f_c=2.45$ GHz. 

\subsection{SIGNAL MODELS}

We adopt the assumptions of \cite{10742896}, where all channels are narrowband line-of-sight (LoS), flat-fading, and in the far-field. There is a single-antenna transmitter (Tx) and $K$ single-antenna Rxs, with no direct path between the Tx and each one of the Rxs. The signal $y_k$ at the $k$-th Rx is given by
\begin{equation}
    y_k=\boldsymbol{h}_k^T\boldsymbol{\Gamma g}s+n_k,
\end{equation}
where $s$ is the transmitted signal, $n_k$ is Additive White Gaussian Noise (AWGN), and $\boldsymbol{h}_k$ and $\boldsymbol{g}$ are the $M\times1$ channels from the RIS to the $k$-th Rx and from the Tx to the RIS. The RIS response is represented by a diagonal matrix containing the RIS reflection coefficients:
\begin{equation}
    \boldsymbol{\Gamma}=\text{diag}[\Gamma(0),\Gamma(1),\ldots,\Gamma(M-1)].
\end{equation}
The element responses are passive ($|\Gamma(m)| \leq 1$ for all $m=0,1,\ldots,M-1$) and are determined by the biasing voltage applied to the $m$-th RIS element. The Tx is assumed to be located at the broadside of the RIS, so the Tx-RIS channel is given by
\begin{equation}
    \boldsymbol{g}=[1,1,\ldots,1] ,
\end{equation}
and the channels from the RIS to each Rx are given by
\begin{equation}
    \boldsymbol{h}(\theta_k) =
    [1,e^{-j\kappa(\theta_k)},e^{-j2\kappa(\theta_k)},\ldots,e^{-j(M-1)\kappa(\theta_k)}],
\end{equation}
where $\theta_k$ is the azimuth angle of the $k$-th Rx from the RIS and $\kappa(\theta)=2\pi(d_xf_\text{c}/c) \sin(\theta)$. We assume the far-field channels remain static during the operation of the RIS, to ensure the integrity of the ML model designed by the initial training data.
%since far-field channels are generally modeled under the planar wavefront assumption, which only depends on the angle of departure or arrival of the signal \cite{9693928}.

The SNR at each Rx $k$ is defined as the ratio of the received signal power divided by the noise power $\sigma_s^2$:
\begin{equation}
    \text{SNR}_k
    =\frac{|E[y_k]|^2}{\sigma_s^2}
    =\frac{|E[\boldsymbol{h}_k^T\boldsymbol{\Gamma g}x]|^2}{\sigma_s^2}
    =\frac{\rho_s |\boldsymbol{h}_k^T\boldsymbol{\Gamma g}|}{\sigma_s^2},
\end{equation}
where $\rho_s$ is the average transmitted symbol power. To evaluate the quality of the radiation patterns generated by the wave-controlled RIS, we define the following SLNR (signal-to-leakage-plus-noise ratio) measure for $K \geq 1$ desired Rx and $L \geq 0$ undesired Rx (``eavesdroppers"): 
\begin{equation}
    \max{\text{SLNR}}=\max\frac{\min_{i\in{1,2,...,K}}{{\rho_s|\boldsymbol{h}_{d,i}^T\boldsymbol{\Gamma g}|^2}}}
    {\max_{j\in{1,2,...,L}}{{\rho_s|\boldsymbol{h}_{e,j}^T\boldsymbol{\Gamma g}|^2}}+\sigma_s^2},
    \label{eq:slnr}
\end{equation}
where each $\boldsymbol{h}_{d,i}$ denotes the channel from the RIS to the $i$-th desired Rx, and $\boldsymbol{h}_{e,j}$ denotes the channel from the RIS to the $j$-th eavesdropper Rx. The independent variable used for optimization is $\boldsymbol{W}$, which determines $\boldsymbol{\Gamma}$, which directly affects the SLNR measure, as seen in~(\ref{eq:slnr}).

\begin{comment}
\begin{enumerate}
    \item Maximize the SNR for a given Rx $k$:
    \begin{equation}
        \max{\text{SNR}_k}=\max{\rho_s|\boldsymbol{h}_k^T\boldsymbol{\Gamma g}|^2}
    \end{equation}
    \item Maximize the SLNR (signal-to-leakage-plus-noise ratio) for $K \geq 1$ desired Rx and $L \geq 0$ undesired (or "eavesdropping") Rx, all at different directions:
    \begin{equation}
        \max{\text{SLNR}}=\max\frac{\min_{i\in{1,2,...,K}}{{\rho_s|\boldsymbol{h}_{d,i}^T\boldsymbol{\Gamma g}|^2}}}
        {\max_{j\in{1,2,...,L}}{{\rho_s|\boldsymbol{h}_{e,j}^T\boldsymbol{\Gamma g}|^2}}+\sigma_s^2}
        \label{eq:slnr}
    \end{equation}
    where each $\boldsymbol{h}_{d,i}$ denotes the channel from the RIS to the $i$-th desired Rx, and $\boldsymbol{h}_{e,j}$ denotes the channel from the RIS to the $j$-th eavesdropper Rx.
\end{enumerate}
\end{comment}

\section{MACHINE LEARNING MODEL DESIGN AND CALIBRATION}
\subsection{BACKGROUND}
%There are numerous factors that separate the ideal RIS model often used in simulations from how an actual RIS behaves in a given physical environment. 
Numerous factors differentiate the ideal RIS model often used in simulations from the actual behavior of an RIS in a physical environment. Starting with the design of the biasing TL, there may be impedance mismatches, losses, and reflections along the path, creating signal losses and distortion \cite{929650}. Since the wave-controlled RIS operates with a wide range of BSWs, there may be frequency-dependent losses along the biasing TL and other nonlinearities caused by skin effects, noise, crosstalk, as well as jitter and phase delays from the waveform generator. There is also a non-negligible voltage drop across the rectifier diode. Depending on the peak detector architecture, its output voltage could be affected by the charging and discharging times of the capacitor. These and other non-ideal effects are difficult to model, and can only be characterized using full-wave simulations or extensive experimental measurements. Furthermore, the reflection coefficients created by the varactor biasing may not be perfectly modeled, as varactor characteristics may vary due to environmental (thermal), fabrication, and electrical conditions \cite{sedra2020microelectronic}, as well as mutual coupling between elements \cite{9319694}. The metasurface response also varies according to the angle of arrival of the incoming signal \cite{7870611}. Lastly, there is the issue of channel estimation for the Tx-RIS and RIS-Rx links required to simulate the performance of the entire system and optimize the resulting beam patterns \cite{9530717, 9328501, 8683663, 9771077}.

\begin{figure}
    \centering
    \includegraphics[width=1.0\linewidth]{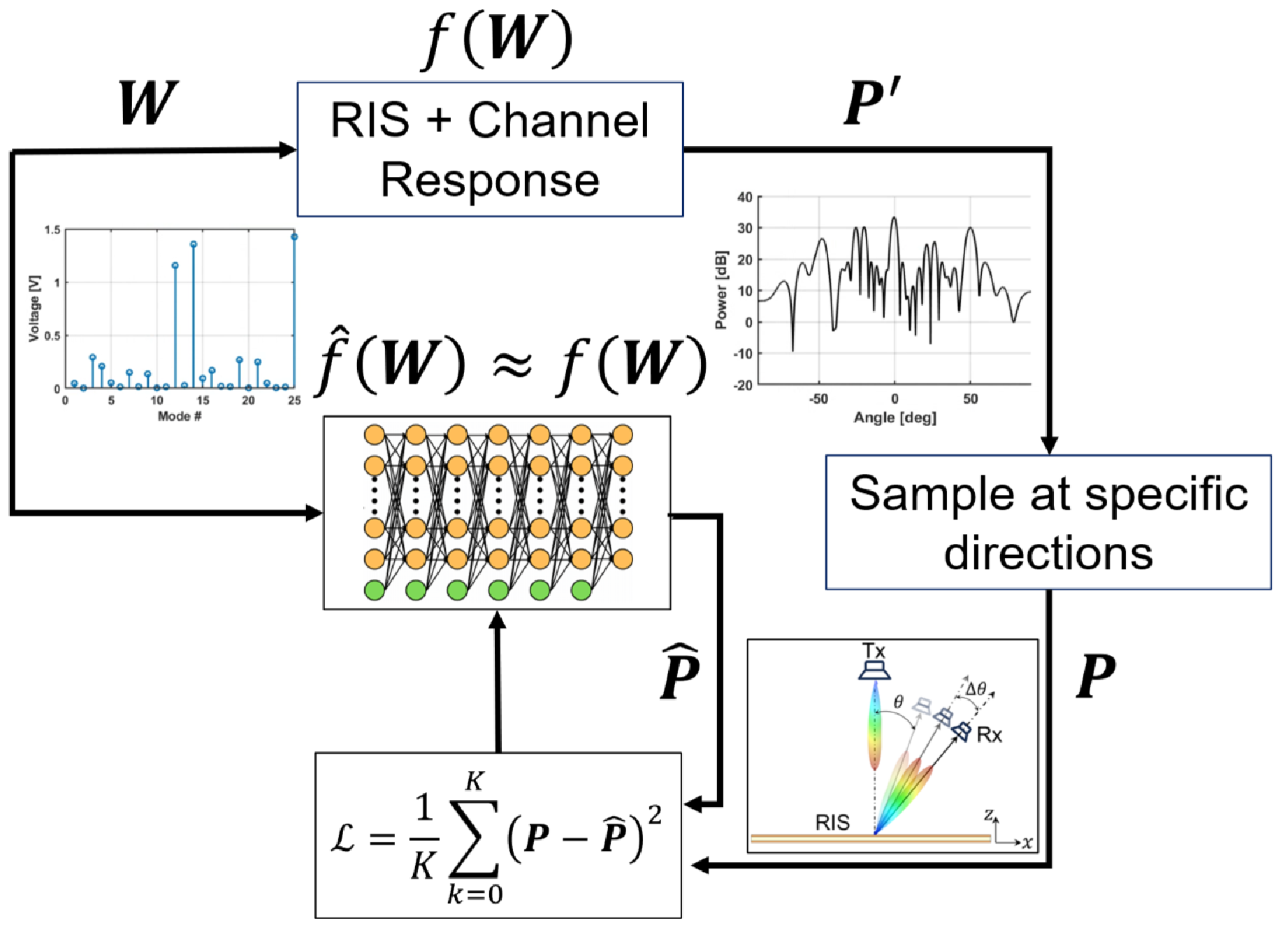}
    \caption{The RIS and channel response are represented by the black box $f(\boldsymbol{W})$ which takes an input set of BSW amplitudes $\boldsymbol{W}$ and outputs a radiation pattern. The radiation pattern $\boldsymbol{P}'$ is sampled at specific locations and stored as $\boldsymbol{P}$. An NN is then trained to estimate the sampled power values $\boldsymbol{\hat{\boldsymbol{P}}}$ from the same data, where the MSE between the actual and estimated power values is used as the loss function $\mathcal{L}$.}
    \label{fig:blackbox}
\end{figure}

Rather than relying on mathematical models that attempt to mimic the RIS and physical environment, an alternative is to use data from the environment in which the RIS is deployed to characterize the behavior of the entire system -- from the input BSW amplitudes to the waveguide and to the resulting radiation patterns. When an RIS is initially deployed, CSI is unavailable and the waveform sampling required to excite the metasurface is unknown. To learn how the system behaves, we can excite the RIS with a combination of BSW amplitudes and sample the power obtained at receivers in different directions of interest. This procedure can be repeated many times to obtain improved knowledge of the input-output relationship of the system, which can in turn be used to design an ML model that models this relationship. The corresponding black box representation of the system and the ML model training procedure can be visualized as in Fig.~\ref{fig:blackbox}. If the ML model is well calibrated, it will provide a sufficiently accurate representation of the nonlinear and complex mapping between the BSW amplitudes and the sampled powers, and can be used to perform offline optimization to obtain the desired beam patterns. Once found, the optimal BSW amplitudes are used to excite the physical RIS and create the actual radiation patterns.

The main problems that must be solved when using this approach include
\begin{enumerate}
    \item What combinations of BSW amplitudes should be used to excite the RIS?
    \item Which ML model provides a sufficiently generalized representation of the system?
\end{enumerate}
The answer to the first question requires some knowledge of the system, such as how large should the maximum voltage swing on the biasing TL be. For this question, for example, we can simulate the ideal scenario where there is no attenuation on the biasing TL, and use these results to verify that the peak voltage generated by each set of BSW amplitudes will not exceed the maximum voltage swing constraint. The answer to the second question requires a deeper analysis of available network architectures and their optimization. Creating a map between the $N$ input BSW amplitudes and the $n_{\text{angles}}$ spatially distributed power values at the receivers is fundamentally a regression problem. Since the map is high-dimensional, nonlinear, and complex, most kernel methods and surrogate function estimators that involve linear transformations and interpolations between training samples will fail. Such approaches also require large matrix operations as more training samples are added \cite{10.1007/978-3-540-45179-2_53, 7547360}. 

An NN, or more specifically, a fully-connected (dense) Multilayer Perceptron (MLP) \cite{mlp} is a better approach for modeling the system, as they achieve universal function approximation by leveraging multiple layers of nonlinear functions with many hidden variables \cite{HORNIK1989359}. The MLP approach takes an input set of $N$ BSW amplitudes, propagates them through multiple hidden layers with nonlinear activation functions, and provides as output a set of $n_{\text{angles}}$ power values from the corresponding radiation pattern, sampled at discrete locations. This is a supervised learning problem, where the network tries to minimize the mean squared error (MSE) between the actual and estimated outputs, while attempting to generalize the complex relationship between the BSW amplitudes and their corresponding radiation patterns. To come up with a good MLP architecture for the given problem, one may choose to manually design an optimal MLP using trial-and-error, which is highly time consuming. An alternative is to use a GA that will create a large pool of MLPs trained using the back-propagation algorithm, evolve them, and converge to an optimal architecture for the given criteria \cite{7731699, 2523}.

The following subsections will address the data processing steps required to use the GA to generate an MLP that will estimate the desired mapping with high accuracy. It is important to note that the procedures discussed in this paper are not only relevant for the specific wave-controlled RIS, rectifier-based multi-user beamforming problem considered here. The proposed procedure and algorithms can be generalized to other RIS modeling and optimization problems, including calibration of a wave-controlled RIS with the sample-and-hold circuit in \cite{10742896}.

\subsection{DATA GENERATION AND PROCESSING}
\begin{algorithm}
    \caption{Dataset Generation}
    \label{alg:dataset}
    \begin{algorithmic}[1]
        \State {$i \gets 0$}
        \While {$i < N_\text{max}$}
            \State {$\boldsymbol{W} \gets [0,0,\ldots,0]^T$}
            \State {$\boldsymbol{W}(n) \gets$ sample from $|\mathcal{N}(0,\sigma_1^2)|$ for $n=1,\ldots,N$.}
            \State {$k \gets$ integer sample from $\mathcal{U}(1,N)$.}
            \State {$\boldsymbol{S} \gets$ unique set of $k$ uniformly chosen integers from 1 to $N$.}
            \State {$\boldsymbol{W}(n) \gets$ sample from $|\mathcal{N}(0,\sigma_2^2)| \; \forall n\in \boldsymbol{S}$.}
            \State {Calculate  maximum absolute value $v_{\text{max}}$ of the resulting BSW superposition over time from $W_0$ and $\boldsymbol{W}$.}
            \If {$v_{\text{max}}$ is out of varactor biasing bounds}
                \State {Go back to line 3.}
            \EndIf
            \State {Sample the power at all $n_{\text{angles}}$ desired directions, store in $\boldsymbol{P}$.}
            \State {Store $\boldsymbol{W}$ and $\boldsymbol{P}$ as $\boldsymbol{W}_i$ and $\boldsymbol{P}_i$.}
            \State {$i \gets i + 1$}
        \EndWhile
    \end{algorithmic}
\end{algorithm}
Simulation of a wave-controlled RIS and corresponding wireless channels was executed using MATLAB with the models defined in Section II. The RIS is modeled as a single-row surface with $M=100$ elements, biased by a superposition of $N=25$ BSWs. The RIS elements are separated by $d_x=20$ mm, while behind the surface the length of the TL path between adjacent rectifiers is $L_\text{p}=131.42$ mm. The effective permittivity of the TL is $\epsilon_\text{eff} = 8.66$, and therefore the slowness factor is $n_\text{slow} = 19.34$. The transmitted signal has carrier frequency $f_{\text{c}}=2.45$ GHz, 
%, making the RIS element separation in terms of wavelengths $\Delta_x \approx 0.1634$. 
and the DC voltage offset is $W_0=4$ V, as this is the minimal voltage required to excite the varactor, and as seen in~(\ref{eqn:wmt1}), the BSWs will always create a higher voltage than $W_0$ at any RIS element location. This allows generation of voltages across the entire allowable range of the varactor. The distance between the beginning of the TL and the first RIS element along the $x$ direction is $L_{\text{left}}=0.5 d_x$. The distance is the same between the last RIS element and the end of the TL. For better control of the voltage over the length $L=99d_x$ used to bias the RIS elements, an additional TL of the same length and geometry as the first biasing TL is connected at its end through a 50 mm connector. Thus, the overall length from the last varactor to the end of the extended TL, normalized by $d_x/L_p$, is
\begin{equation}
    L_{\text{right}} = 0.5d_x+50\text{mm}\cdot\frac{d_x}{131.42\text{mm}}+100d_x\approx100.88 d_x
\end{equation}
% \begin{equation}
%     M_r = 0.5+\frac{50\text{mm}}{131.42\text{mm}}+101 \approx % 101.88.
% \end{equation}
This extended TL results in a fundamental BSW frequency of $f_{\text{b}}=1.93$ MHz.

To enable good coverage of the 25-dimensional input space, a dataset was generated with $N_\text{max}=$ 100,000 pairs of data arrays, each consisting of a random $\boldsymbol{W}$ array of voltages as the input, and samples from the resulting radiation pattern  vector $\boldsymbol{P}$ in dB as outputs when the RIS waveguide is excited by the corresponding BSWs. 
%In the actual physical scenario, there would need to be a receiver in each location of interest. Receivers should preferably be spaced by less than a beamwidth for areas where wider coverage is required, for precise beamforming and nullforming. By separating the sample points by less than a beamwidth, we minimize the variance between measurements and obtain a close approximation of the entire radiation pattern. 
The azimuth angles corresponding to the sampled points on the radiation pattern are in the interval [$-60\degree,60\degree]$. The simulated 3 dB beamwidth for the given $M=100$ element RIS is approximately $1.7\degree$, so we used discrete steps of $1.5\degree$ spaced uniformly between $[-60\degree, 60\degree]$, resulting in power samples at $n_{\text{angles}}=81$ different angles.

To reduce the likelihood of producing chaotic radiation patterns caused by fully randomized inputs which will negatively affect the learning by the MLP ~\cite{taghvaee2021radiation}, the following methodology is proposed. Since the MLP will later be used to estimate the model for optimization by SA, which randomly perturbs all the BSW amplitudes in the input vector to converge towards an optimal SLNR measure, a small random Gaussian voltage with a standard deviation $\sigma_1=0.008$ V whose absolute value is taken, is added to all elements of an initially empty $\boldsymbol{W}$. A randomly selected subset of the BSW amplitudes is then excited by Gaussian-distributed voltages with a much larger standard deviation $\sigma_2=0.8$ V, also converted to absolute values, to put more emphasis on those BSWs and create more unique radiation patterns. Note that only positive voltage amplitudes are used to reduce the dataset complexity by truncating the Gaussian distributions by taking the absolute values of their samples. This ensures that only the amplitudes of the BSWs are modulated, rather than both the amplitudes and the phases. The entire procedure to create the training dataset is described in Algorithm~\ref{alg:dataset}. Since we use back-propagation and gradient descent, it is beneficial to normalize the data to improve convergence \cite{589532}. In particular, we scale each of the 100,000 $\boldsymbol{W}$ vectors to be within the range $[0, 1]$, and we scale the resulting power values in $\boldsymbol{P}$ to be in the range $[-1, 1]$.

\subsection{NEURAL NETWORK OPTIMIZATION USING A GENETIC ALGORITHM}

\begin{comment}
    
\begin{table}
\caption{Optimal Neural Network Architecture.}
\label{table:mlp}
\centering
\begin{tabular}{|l|l|l|}
\hline
\multicolumn{1}{|c|}{Hidden Layer} & 
\multicolumn{1}{c|}{Number of Nodes} & 
\multicolumn{1}{c|}{Activation Function} \\ 
\multicolumn{1}{|c|}{Index} & 
\multicolumn{1}{c|}{} & 
\multicolumn{1}{c|}{} \\ \hline
1 & 1024 & PReLU \\ \hline
2 & 64 & tanh \\ \hline
3 & 512 & PReLU \\ \hline
4 & 2048 & PReLU \\ \hline
5 & 1024 & ReLU \\ \hline
\end{tabular}
\end{table}

\end{comment}

An MLP generally includes multiple hyperparameters that can be tuned to optimize its architecture. These include the number of hidden layers, the number of nodes per layer, the type of activation function used, the number of epochs (learning iterations over the entire training set), and the batch size (number of samples used before each back-propagation step) \cite{smith2018disciplinedapproachneuralnetwork}. One method to optimize these hyperparameters is by using a GA, which is a global search algorithm based on an evolutionary model that is used to solve difficult non-convex optimization problems \cite{forrest1996genetic}. It works by storing binary strings that contain the information about individuals in a population and evolving them by simulating the natural selection process. The GA approach is generally implemented with the following steps: Generate an initial population consisting of individuals created randomly, with each individual represented by a string of bits as a candidate solution for the problem of interest. Using the principle of natural selection, individuals may reproduce (copy themselves) into the next generation depending on their fitness, which is determined by the desired objective function (selection process). Individuals may also mutate (random bit flips) or crossover (exchange of bits between two different individuals) to create new individuals in the next generation (children). As generations evolve over multiple iterations, the population will improve, as the individuals represent better solutions to the given problem.

The general GA described above can customized for MLP architecture optimization. Since the number of hyperparameters in an MLP is variable, % the GA should be customized to optimize the MLP architecture., 
crossover and mutation on individual MLPs with different numbers of hyperparameters must be carefully defined. The authors in \cite{DOMASHOVA2021263} propose a GA that finds an optimal architecture for an MLP for a given classification task, but a similar approach can also be applied for regression tasks, as discussed in this section. The GA is implemented as following:
\begin{itemize}
    \item \textbf{Initial population} -- We start with a collection of MLP realizations or ``individuals'' with randomly chosen numbers of epochs, batch sizes, numbers of hidden layers, numbers of neurons in each layer, and types of activation functions.
    \item \textbf{Objective} -- The fitness of each MLP is defined as the validation loss from training it on the same dataset containing the RIS BSW amplitudes and corresponding radiation patterns, by evaluating the MSE between the predicted radiation patterns and the actual radiation patterns for a given BSW amplitudes array.
    \item \textbf{Parent selection} -- Parents are randomly selected individuals from the population, and their children are reproductions of themselves with probabilities for crossover and mutation.
    \item \textbf{Crossover} -- The number of epochs, batch size, and number of hidden layers of the child are each chosen to be equal to the corresponding value of one of the parents with equal probability between the two parents. For each hidden layer, the child will inherit randomly the activation function and the number of nodes from the corresponding layer of either parent. If the number of hidden layers of the parents are different, then there may be a case where the child inherits the larger number of layers, which is best explained by an example. Suppose that Parent A has 5 hidden layers, Parent B has 3 hidden layers, and the child has 5 hidden layers. Hidden layers 1-3 of the child will have randomly chosen parameters between layers 1-3 of each parent. For layers 4-5, the parameters will be randomly chosen between layers 4-5 of Parent A, or layer 3 of Parent B.
    \item \textbf{Mutation} -- Only occurs when a child has a larger number of hidden layers than one of the parents. This involves random shuffling of the number of nodes and activation functions between all hidden layers of the child.
    \item \textbf{Evolution} -- Only the fittest $k$ individuals from the current population (the original MLPs and the children) move to the next generation, where individuals reproduce again and the cycle repeats. The number $k$ decreases during each evolution step.
\end{itemize}
The algorithm converges after the populations evolve all the way to a single MLP with the lowest validation loss.

In our simulations, the GA was implemented with randomly chosen parameters for each MLP from the following sets:
\begin{itemize}
    \item Number of epochs: \{80, 81, \ldots, 200\}.
    \item Batch sizes: \{32, 64, 128, 256, 512, 1024\}.
    \item Number of hidden layers: \{1, 2, \ldots, 6\}.
    \item Number of nodes per hidden layer: \{64, 128, 256, 512, 1024, 2048\}.
    \item Activation functions \cite{8407425}: \{ReLU, PReLU, Sigmoid, Tanh\}.
\end{itemize}
The initial population has $R$ randomly generated individuals. Then, $\frac{R}{2}$ randomly selected pairs of parents are chosen with replacement to have 2 children each. From both the original population and the new children, a population of $k=\frac{R}{2}$ individuals with the lowest validation loss is created. This process is repeated until there is only a single individual left. In our simulations, we chose $R$ to be 64 as a good tradeoff between the number of solutions explored versus the time to convergence. A total of 190 MLPs were trained, and the population evolved over 7 generations until the final MLP was determined.

\begin{figure}
    \centering
    \includegraphics[width=1.0\linewidth]{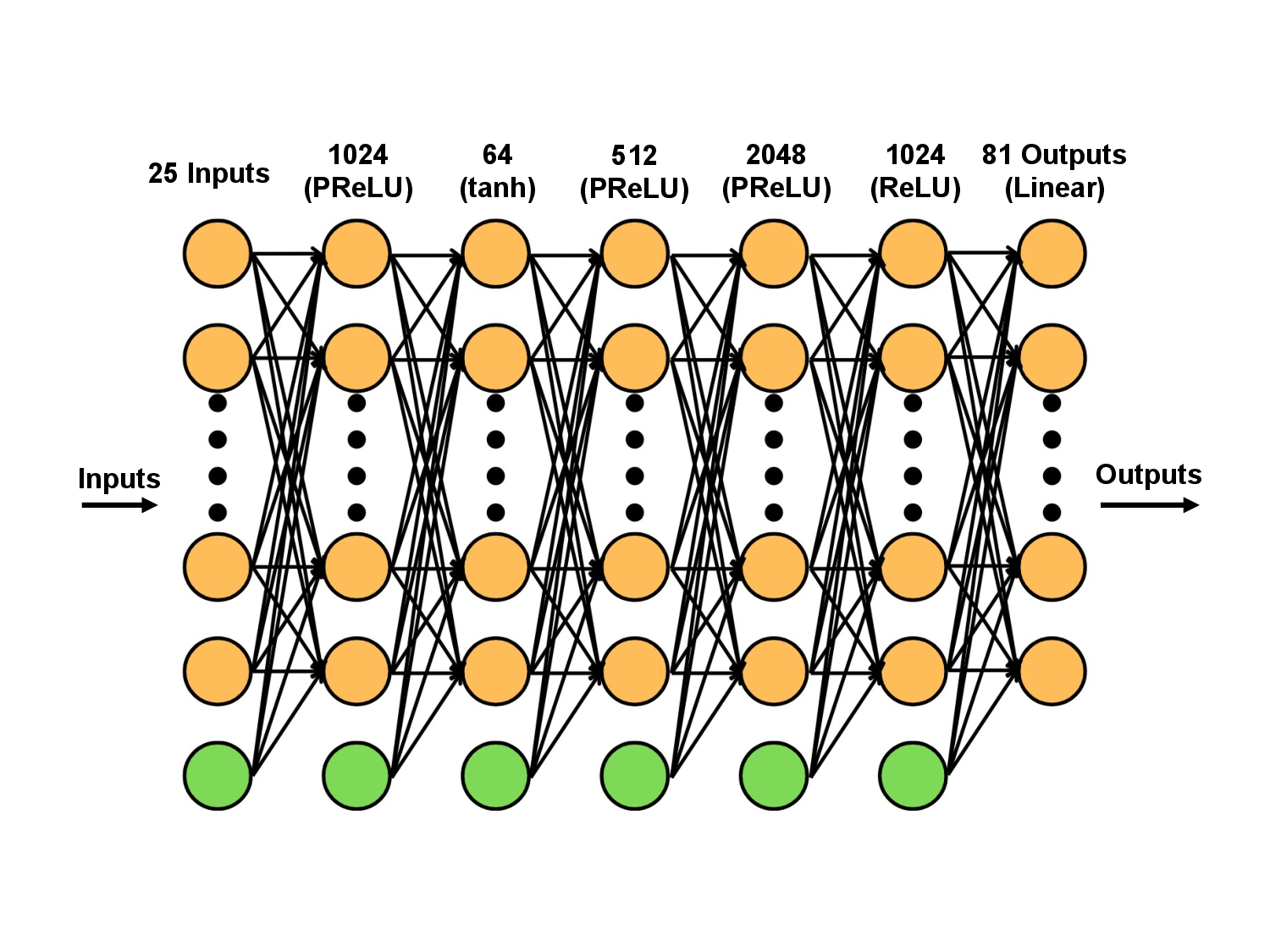}
    \caption{Illustration of the proposed NN, including bias nodes (green) in the input layer and each hidden layer. The number and type of activation functions in each layer are given above each layer. The black arrows correspond to the weights between neurons. Inputs flow from the left, through the neurons in the hidden layers of the network, to the outputs on the right.}
    \label{fig:mlp}
\end{figure}

The MLPs were implemented using TensorFlow Keras \cite{tensorflow2015-whitepaper}. All MLPs were trained with an L2 regularization penalty of $\SI{5e-7}{}$ to improve generalization \cite{Shi2019}, using the Adam Optimizer \cite{kingma2017adammethodstochasticoptimization} and the MSE loss between the actual and predicted scaled power values. The learning rate is reduced by half if the validation loss stops improving after 20 consecutive epochs, and the training is terminated if the validation loss does not improve after 30 consecutive epochs. The training set consisted of the first 90,000 samples from the dataset and the validation set consisted of the last 10,000 samples. The fitness of each MLP was determined by its validation loss.

The final MLP architecture that resulted from the GA optimization has 5 hidden layers with nodes and activation functions as illustrated in %Table~\ref{table:mlp} and 
Fig.~\ref{fig:mlp}. The MLP has a batch size of 128 and goes through 147 training epochs. The final MSE training loss was $\SI{7.4851e-4}{}$ and the final MSE validation loss was $\SI{8.4779e-4}{}$.

\section{OFFLINE OPTIMIZATION USING SIMULATED ANNEALING}
\subsection{MACHINE LEARNING MODEL PERFORMANCE}

\begin{table*}
    \centering
    \begin{tabular}{|c|c|c|c|l||c|c|c|c||c|c|c|c|}
    \hline
    \multicolumn{5}{|c||}{} & 
    \multicolumn{4}{c||}{\textbf{NN System}} & 
    \multicolumn{4}{c|}{\textbf{Simulated System}} \\ \hline
     Beam 1 & Beam 2 & Null 1 & Null 2 & & Avg & Min & Max & Std. dev & Avg & Min & Max & Std. dev \\ \hline
     \multirow{2}*{-40.5} & \multirow{2}*{-} & \multirow{2}*{-} & \multirow{2}*{-} & Real & 27.8578 & 27.2220 & 28.6893 & 0.5420 & 31.7187 & 30.5748 & 33.6562 & 0.8947 \\ \cline{5-13}
     & & & & Estimated & 28.8694 & 27.8159 & 30.1901 & 0.8664 & - & - & - & - \\ \hline
     \multirow{2}*{25.5} & \multirow{2}*{-} & \multirow{2}*{-} & \multirow{2}*{-} & Real & 32.7547 & 32.3752 & 33.9274 & 0.5156 & 33.6853 & 32.5867 & 34.3495 & 0.6721 \\ \cline{5-13}
     & & & & Estimated & 36.1271 & 35.2647 & 36.5626 & 0.3988 & - & - & - & - \\ \hline
     \multirow{2}*{-30} & \multirow{2}*{-} & \multirow{2}*{-49.5} & \multirow{2}*{-} & Real & 27.7722 & 24.0745 & 31.4893 & 3.3380 & 29.9627 & 25.6659 & 33.0271 & 2.1174 \\ \cline{5-13}
     & & & & Estimated & 29.0995 & 25.0857 & 32.5343 & 3.2351 & - & - & - & - \\ \hline
     \multirow{2}*{-19.5} & \multirow{2}*{-49.5} & \multirow{2}*{-} & \multirow{2}*{-} & Real & 26.5683 & 22.9833 & 28.4797 & 2.3155 & 27.6214 & 25.8481 & 28.6254 & 0.9598 \\ \cline{5-13}
     & & & & Estimated & 27.5753 & 24.1129 & 29.6426 & 2.2093 & - & - & - & - \\ \hline
     \multirow{2}*{-19.5} & \multirow{2}*{49.5} & \multirow{2}*{10.5} & \multirow{2}*{-40.5} & Real & 19.3474 & 14.6896 & 24.4148 & 3.4994 & 21.4651 & 17.0254 & 24.7522 & 2.2578 \\ \cline{5-13}
     & & & & Estimated & 20.7587 & 15.0296 & 24.4111 & 3.9277 & - & - & - & - \\ \hline
     \multirow{2}*{30} & \multirow{2}*{-} & \multirow{2}*{0} & \multirow{2}*{-} & Real & 4.3228 & 2.1095 & 5.6858 & 1.0251 & 6.4071 & 2.3809 & 9.7270 & 2.5564 \\ \cline{5-13}
     & & & & Estimated & 6.0771 & 5.1976 & 7.5983 & 0.6638 & - & - & - & - \\ \hline
    \end{tabular}
    \caption{SLNR [dB] After Convergence of Simulated Annealing for the NN System and the Simulated System.}
    \label{table:sa_1}
\end{table*}

\begin{table*}
    \centering
    \begin{tabular}{|c|c|c|c|l||c|c|c|c||c|c|c|c|}
    \hline
    \multicolumn{5}{|c||}{} & 
    \multicolumn{4}{c||}{\textbf{NN System}} & 
    \multicolumn{4}{c|}{\textbf{Simulated System}} \\ \hline
     Beam 1 & Beam 2 & Null 1 & Null 2 & & Avg & Min & Max & Std. dev & Avg & Min & Max & Std. dev \\ \hline
     \multirow{2}*{-35} & \multirow{2}*{-} & \multirow{2}*{-50} & \multirow{2}*{-} & Real & 25.4248 & 11.0920 & 30.4218 & 5.8103 & 29.2951 & 26.4624 & 31.4586 & 1.6349 \\ \cline{5-13}
     & & & & Estimated & 25.7507 & 13.4766 & 30.3057 & 5.0270 & - & - & - & - \\ \hline
     \multirow{2}*{-20} & \multirow{2}*{50} & \multirow{2}*{10} & \multirow{2}*{-40} & Real & 5.7627 & 2.1418 & 12.4138 & 3.4949 & 24.4674 & 20.9093 & 27.9945 & 2.2030 \\ \cline{5-13}
     & & & & Estimated & 14.0304 & 11.2299 & 18.1563 & 2.3760 & - & - & - & - \\ \hline
    \end{tabular}
    \caption{SLNR [dB] After Convergence of Simulated Annealing for the NN System and the Simulated System for Arbitrary Directions.}
    \label{table:sa_2}
\end{table*}

The NN models the system as the mapping $\mathbb{R}^{N} \rightarrow \mathbb{R}^{n_{\text{angles}}}$ denoted by the nonlinear function $\hat{f}(\boldsymbol{W})= \hat{\boldsymbol{P}}$, where it is desired that $\hat{\boldsymbol{P}} \approx \boldsymbol{P}$, as illustrated in Fig.~\ref{fig:blackbox}. We are interested in offline optimization of the BSW amplitudes to configure the RIS to produce a desired radiation pattern. For this optimization, SA \cite{traveling-salesman-thermo} is a reasonable choice, given our prior success using it to create SLNR-optimized radiation patterns for the wave-controlled RIS implemented with the rectifiers \cite{10742896}. In this case, the NN can be used to estimate the powers sampled at specific directions after each update of $\boldsymbol{W}$ as the SA converges towards improved SLNR values. Once the algorithm converges to a particular set of optimal BSW amplitudes, these amplitudes can be used to reconfigure the physical RIS and generate the optimal radiation pattern.
% The NN takes BSW amplitudes $\boldsymbol{W}$ as inputs and estimates their corresponding sampled radiation patterns $\hat{\boldsymbol{P}}$. Using another optimization algorithm, such as Simulated Annealing, we can optimize the BSW amplitudes such that they produce a desired radiation pattern based on a defined cost function (e.g., maximize SNR or SLNR), using the estimations from $\hat{f}(\boldsymbol{W})$. This enables offline optimization of beampatterns, rather than relying on online feedback from the receivers. When the optimization algorithm converges to an optimal set of BSW amplitudes, this set can be used as the input to the real RIS to instantly generate the optimal radiation pattern.

We used a similar SA algorithm as in \cite{10742896} to determine the BSW amplitudes that optimize the SLNR, as outlined in Algorithm~\ref{alg:simulated_annealing}. This is a stochastic optimization that randomly perturbs the set of BSW amplitudes as it searches for improved SLNR. The algorithm starts at a ``high energy'' state corresponding to a maximum temperature parameter $T$, in which it is more likely to explore different BSW amplitudes even if they do not necessarily produce better SLNR values. This ultimately helps in escaping local minima. As the algorithm converges over multiple iterations, it moves to a ``lower energy'' state with a low value for $T$, where it is less likely to explore solutions with lower SLNR. The likelihood of moving from the current set of BSW amplitudes $\boldsymbol{W}$ to the next one $\boldsymbol{W}_{\text{new}}$ is given according to a probability measure
\begin{equation}
    p=\begin{cases}
         1 & \text{SLNR}_{\text{new}} > \text{SLNR}_\text{current} \\
         e^{\left(-\frac{\text{SLNR}-\text{SLNR}_{\text{new}}}{k_{\text{c}} T}\right)} & \text{SLNR}_{\text{new}} \leq \text{SLNR}_{\text{current}},
    \end{cases}
    \label{eq:simulated_annealing_prob}
\end{equation}
which depends on the SLNR corresponding to $\boldsymbol{W}$, the new SLNR corresponding to $\boldsymbol{W}_{\text{new}}$, the temperature $T$, and a cooling factor constant $k_{\text{c}}$. The advantage of the SA algorithm is that it is not computationally intensive -- the random perturbations are simple addition and subtraction operations, and the SLNR values are calculated from the forward pass of the BSW amplitudes through the neural network during inference. The algorithm is initialized with $\boldsymbol{W}=[0,0,...,0]^T$.

\begin{comment}
After each perturbation in $\boldsymbol{W}$ which yields $\boldsymbol{W}_{\text{new}}$ in line 13 of the algorithm, we measure the new SLNR [dB] value estimated by the MLP. We define the update rule for simulated annealing whether to use $\boldsymbol{W}_{\text{new}}$ as the baseline for the next perturbation according to a probability measure
\begin{equation}
    p=\begin{cases}
         1 & \text{SLNR}_{\text{new}} > \text{SLNR}_\text{current} \\
         e^{\left(-\frac{\text{SLNR}-\text{SLNR}_{\text{new}}}{k_{\text{c}} T}\right)} & \text{SLNR}_{\text{new}} \leq \text{SLNR}_{\text{current}}.
    \end{cases}
    \label{eq:simulated_annealing_prob}
\end{equation}
\end{comment}

We compare the performance of SA executed offline by inference using the MLP for $\boldsymbol{W}$ during each iteration estimate the power values, against the performance of SA executed in the deterministic simulation, where $\boldsymbol{W}$ excites the RIS and the exact simulated power values at the Rx are calculated at each iteration. The first system will be referred to as the ``NN System'' and the second system will be referred to as the ``Simulated System.'' The SA algorithms in both cases have the same hyperparameters:
\begin{itemize}
    \item Cooling factor: $k_{\text{c}} = 0.002$.
    \item Maximum number of iterations: $i_{\text{max}} = 2000$.
    \item Random perturbations $\epsilon$ for each BSW amplitude chosen from $\epsilon \sim \mathcal{N}(0,1)$.
    \item Scaling factor for random steps: $\lambda = 0.015$.
    \item Maximum number of iterations between current SLNR and previous best SLNR: $i_\text{rst}=200$.
    \item Temperature scaling factor: $T_\text{scale}=100$.
\end{itemize}

The major difference between the two is that for the NN system, only positive BSW amplitudes are used because the network was trained only on positive amplitudes, whereas for the simulated system, both positive and negative amplitudes are used, to allow another degree of freedom that the NN does not have. This provides a fair comparison between the beam patterns achievable by each system, without limiting the simulated system to the same constraint that the NN system has. We studied multiple scenarios with single beams, multiple beams, and combinations of beams and nulls, and evaluated the performance of each system in terms of SLNR after convergence, assuming a noise variance of $\sigma_{\text{s}}^2=1$ in~(\ref{eq:slnr}). Table~\ref{table:sa_1} compares the SA performance using feedback from the NN versus using direct Rx feedback from the simulated setup where the RIS response and channel models are exact. The final results are divided into two categories: ``Estimated,'' which reports the final SLNR value that SA in the NN system converged to, and ``Real,'' which is the SLNR found by the deterministic simulation using the optimal $\boldsymbol{W}$ arrays that SA in each system converged to.% The ``Real'' SLNR is determined from the final BSW amplitude vector $\boldsymbol{W}$ that SA using the NN converged to, and the final $\boldsymbol{W}$ obtained from SA using direct feedback from the deterministic simulation.
% The MLP in each scenario has two rows: Simulated SLNR (the SLNR after the optimal set of BSW amplitudes was used in the simulation) and Estimated SLNR (the SLNR that the MLP estimates from the optimal set of BSW amplitudes). Each scenario was simulated 10 times and summarized in Table~\ref{table:sa_1}.
% Note that in the last case (beam at $30\degree$ and null at $0\degree$), the algorithm will try to suppress the power in the normal incidence to RIS, by treating $0\degree$ as an undesired / eavesdropper direction in the SLNR cost function. It may not be able to create a real ``null.''

\begin{algorithm}[!t]\caption{Simulated Annealing with Neural Network}\label{alg:simulated_annealing}
    \begin{algorithmic}[1]
        \State $\boldsymbol{W} \gets [0,0,\ldots,0]^T$
        \State $\boldsymbol{W}_{\text{best}} \gets \boldsymbol{W}$
        \State $i_{\text{best}} \gets 0$
        \State $\text{SLNR}_{\text{current}}$ $\gets$ SLNR calculated from (\ref{eq:slnr}) by forward-propagating $\boldsymbol{W}$ through the NN.
        \State $\text{SLNR}_{\text{best}} \gets \text{SLNR}_{\text{current}}$
        \While {$i < i_{\text{max}}$}
            \If {$i-i_{\text{best}}\geq i_\text{rst}$}
                \State $\boldsymbol{W} \gets \boldsymbol{W}_{\text{best}}$
                \State $i_{\text{best}} \gets i$
                \State $\text{SLNR}_{\text{current}} \gets \text{SLNR}_{\text{best}}$
            \EndIf
            \State $T \gets T_\text{scale}\left(1-\frac{i}{i_{\text{max}}}\right)$
            \State $\boldsymbol{W}_{\text{new}}(n)\! \gets \! \left| \boldsymbol{W}(n) \! + \! \lambda \epsilon \right|,\; \ n= 1,2,\ldots,N$
            \State $\text{SLNR}_{\text{new}}$ $\gets$ SLNR calculated from (\ref{eq:slnr}) by forward-propagating $\boldsymbol{W}_{\text{new}}$ through the NN.
            \If{$\text{SLNR}_{\text{new}} > \text{SLNR}_{\text{best}}$}
               \State $\text{SLNR}_{\text{current}} \gets \text{SLNR}_{\text{new}}$
                \State $\text{SLNR}_{\text{best}} \gets \text{SLNR}_{\text{new}}$
                \State $\boldsymbol{W} \gets \boldsymbol{W}_{\text{new}}$
                \State $\boldsymbol{W}_{\text{best}} \gets \boldsymbol{W}_\text{new}$
                \State $i_{\text{best}} \gets i$
            \Else
                \State Calculate $p$ using~(\ref{eq:simulated_annealing_prob})
                \If{$p \geq$ rand(1)}
                    \State $\boldsymbol{W} \gets \boldsymbol{W}_{\text{new}}$
                    \State $\text{SLNR}_{\text{current}} \gets \text{SLNR}_{\text{new}}$
                \EndIf
            \EndIf
        \EndWhile
    \end{algorithmic}
\end{algorithm}

We note that the simulated average SLNR in all cases is quite close between the offline optimization using the NN and the online optimization from the simulation, with the worst matching having around 4 dB of difference between the average SLNRs of the two. This indicates that the MLP was able to generalize the true mapping between BSW amplitudes and radiation patterns to high accuracy. Even for cases where the objective function includes forming deep nulls, which are sensitive to small changes in the reflection coefficients, the MLP nearly matches the performance of the simulated scenario.

Note that the cases illustrated in Table~\ref{table:sa_1} only involve angles on which the MLP was trained. It would also be useful to generalize the algorithm to create beams and nulls in any arbitrary direction within the sampled range.

\subsection{ANGLE INTERPOLATION}
When uniformly sampling the radiation pattern to create the initial dataset, we used the 3 dB beamwidth as the criterion for maximum angular separation between any two sampled Rx directions to reconstruct the radiation pattern. As stated previously, this helps with beamforming and nullforming at Rx directions between any two neighboring sampled directions, because attempting to change the power directed towards one direction will significantly affect the power directed toward its nearby directions within a beamwidth. We can estimate the power values at Rx directions between the discrete samples through linear interpolation, as follows.% This allows to create some estimations regarding the powers directed between the discrete samples, as follows.

We define the floating-point index of each angle of interest $\theta^*$ as
\begin{equation}
    i=\frac{\theta^*-\theta_{\text{min}}}{\theta_{\text{max}}-\theta_{\text{min}}}\times(n_{\text{angles}}-1),
\end{equation}
where $\theta_{\text{min}}$ is the smallest (most negative) angle sampled in the training dataset (in this case, $-60\degree$) and $\theta_{\text{max}}$ is the largest (most positive) angle in the training dataset ($60\degree$). Since the powers are sampled at discrete angles represented by integer indices in the interval $[0,n_{\text{angles}}-1]$, we calculate the fractional part of $i$ as
\begin{equation}
    \Delta_{i} = i-\lfloor i \rfloor.
\end{equation}
Finally, we define $P_{-}$ and $P_{+}$ as the powers sampled at the angles corresponding to $\lfloor i \rfloor$ and $\lceil i \rceil$, respectively. When evaluating the SLNR measure, we calculate the power at interpolated directions using the convex combination
\begin{equation}
    P=(1-\Delta_{i})P_{-}+\Delta_{i}P_{+}
    \label{eq:p_int}
\end{equation}
%following the intuition that if a beam is created between two discrete directions, the power corresponding to the two angles should be maximized, and the closer the beam is to one angle, more emphasis should be placed on that same angle. A similar statement can be made regarding nulls, where the power directed towards both angles should be suppressed, with more emphasis on suppressing the power at the angle closer to the direction of interest.

We simulated two cases involving beam- and nullforming using the same SA algorithm as before, except that the power calculated in each direction by the MLP in the NN system is defined as in~(\ref{eq:p_int}). The simulated system is allowed to sample the power in any given direction rather than being confined to a discrete grid. The results are summarized in Table~\ref{table:sa_2} averaged over 10 simulations. We note that the MLP performance is degraded in this case, although it is able to create both beams and nulls. This degradation can be seen both by the low mean SLNRs and the high standard deviations. The primary reason for this degradation is that the SLNR objective function becomes too complicated for SA to maximize when we start with an empty $\boldsymbol{W}$ vector, since it must account for more directions in both beamforming and nullforming. In the next section we consider the use of a lookup table to improve algorithm convergence, by storing optimal $\boldsymbol{W}$ vectors in memory and using them as initialization for SA for more complex cases. 
\section{ADAPTIVE OPTIMIZATION USING A LOOKUP TABLE}

A lookup table can be used to store BSW amplitude arrays that correspond to radiation patterns that were already optimized for different cases. This serves two main purposes: First, the optimal arrays can be used to configure the RIS instantly to create desired radiation patterns that have already been generated in the past, thus requiring fewer executions of the SA algorithm. Second, the optimal arrays can be used to optimize related, but more complex radiation patterns, by being the initial arrays used in the SA algorithm. For example, if there was a requirement in the past to generate a beam at direction $\theta_1$, the same BSW amplitudes that were designed to create this beam can be used as the baseline to produce a radiation pattern with a beam at $\theta_1$ and another beam at $\theta_2$. This approach greatly helps with convergence as multiple beams and/or multiple nulls are desired, as demonstrated here.

% A lookup table can be used to store BSW amplitudes that correspond to radiation patterns that were optimized for different cases, to be used as an initialization for optimizations involving related but more complex scenarios. Since the ML model was trained on a large dataset, it is likely that this dataset will already contain BSW amplitude arrays that correspond to radiation patterns of interest, such as those featuring strong beams or deep nulls in specific directions. These can be used as bases for optimization and to adaptively create the lookup table. As the lookup table stores more optimal BSW amplitude arrays, the system behavior is learned adaptively over time and fewer optimizations will be required.

We propose the following approach to construct the lookup table, as outlined in Algorithm~\ref{alg:lookup}, where $\boldsymbol{\theta}_{\text{beam}}$ and $\boldsymbol{\theta}_{\text{null}}$ are arrays respectively containing the $K$ and $L$ directions in which beams and nulls are to be created. Assume for example that we wish to create a single beam in a specific direction. Given the large dataset used for training, which is the collection of random $\boldsymbol{W}$ vectors and their corresponding sampled power values, we find the BSW amplitude input $\boldsymbol{W}$ that corresponds to the highest power ($P_{\text{max}}$) in the same direction as the desired beam, and we use $\boldsymbol{W}$ as the initialization for SA. After SA converges, we store the resulting $\boldsymbol{W}$ in the lookup table with its corresponding beam direction. To create beams in two directions, we repeat the same procedure -- find the strongest beam between the two in the lookup table and use this result as the basis to design the second beam using SA. The same procedure can be generalized for larger numbers of beams as well. When forming both beams and nulls, we repeat the same procedure as above to create all the beams first, and then create all the nulls in the last optimization steps. The table may contain an entry of the form: \\
\textbf{Beams:} \{$\theta_\text{b1},\theta_\text{b2}$,\ldots\}; \textbf{Nulls}: \{$\theta_\text{n1},\theta_\text{n2}$,\ldots\}; \textbf{SLNR}: $y$; $\boldsymbol{W}.$

\begin{figure*}[!ht]
    \centering
    \begin{subfigure}[t]{0.24\textwidth}
        \includegraphics[width=\textwidth]{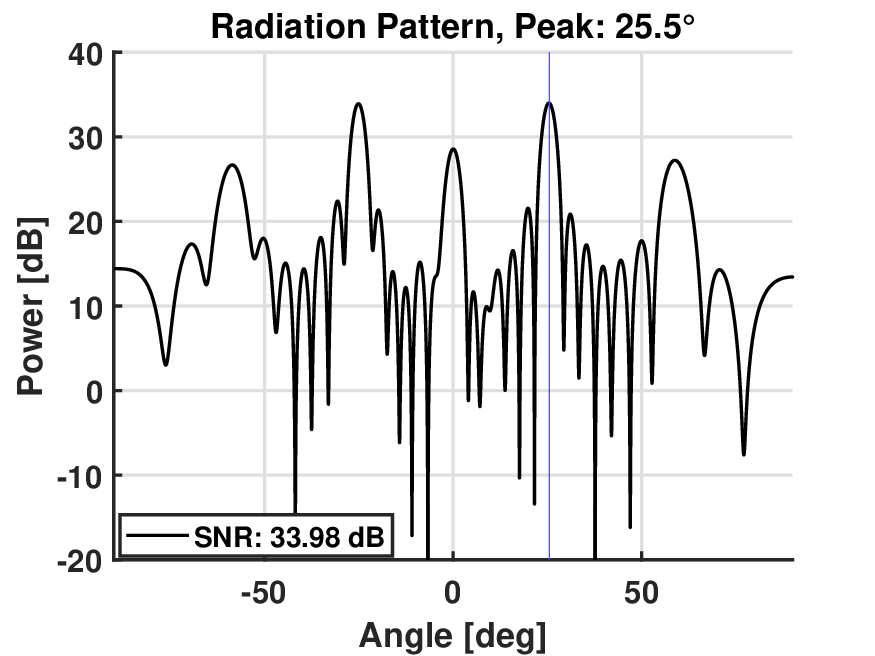}
        \caption{Single beam at $25.5\degree$. SNR: 33.98 dB.}
        \label{fig:rad_patterns_a}
    \end{subfigure}
    \hfill
    \begin{subfigure}[t]{0.24\textwidth}
        \includegraphics[width=\textwidth]{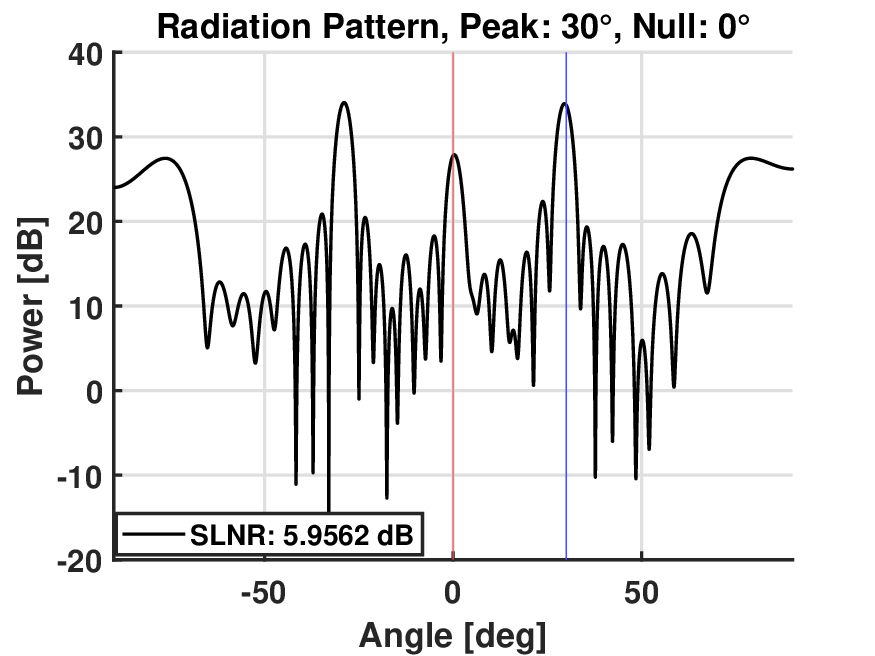}
        \caption{Single beam at $30.0\degree$, suppressing power at $0\degree$. SNR at $30\degree$: 33.7412 dB; SLNR: 5.9562 dB.}
        \label{fig:rad_patterns_b}
    \end{subfigure}
    \hfill
    \begin{subfigure}[t]{0.24\textwidth}
        \includegraphics[width=\textwidth]{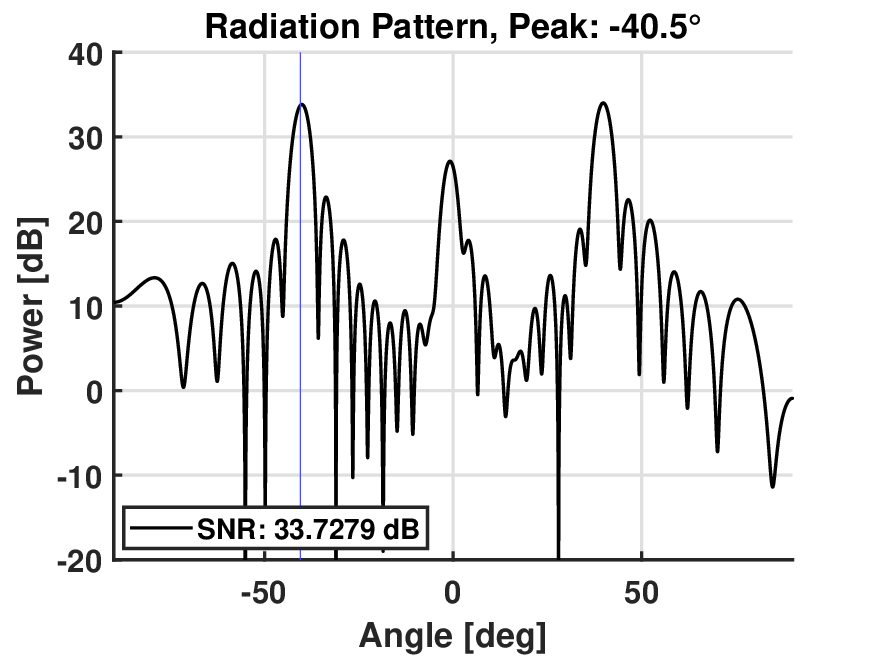}
        \caption{Single beam at $-40.5\degree$. SNR: 33.7279 dB.}
        \label{fig:rad_patterns_c}
    \end{subfigure}
    \hfill
    \begin{subfigure}[t]{0.24\textwidth}
        \includegraphics[width=\textwidth]{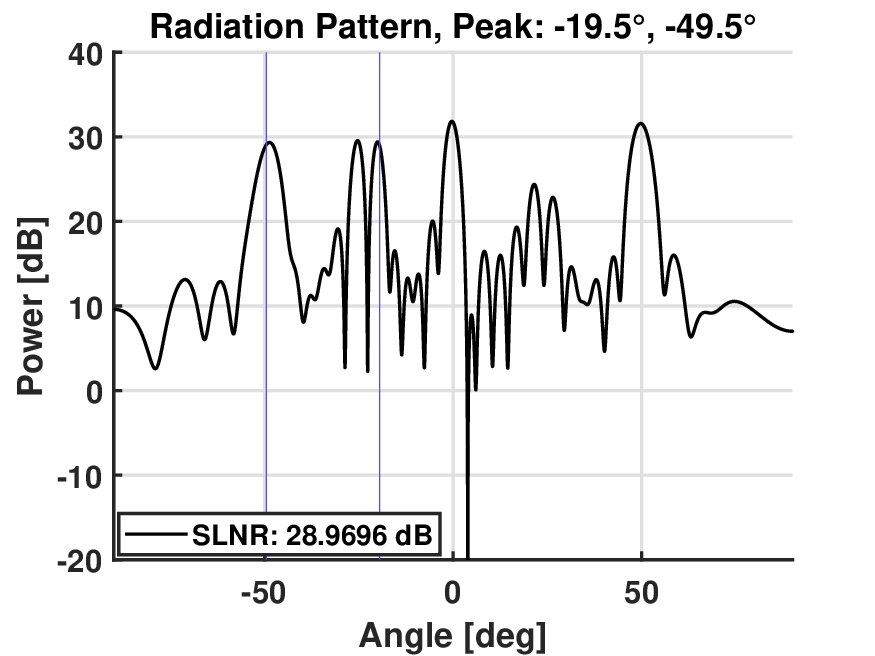}
        \caption{Beams at $-19.5\degree$ and $-49.5\degree$. SLNR: 28.9696 dB.}
        \label{fig:rad_patterns_d}
    \end{subfigure}
    \hfill
    \begin{subfigure}[t]{0.24\textwidth}
        \includegraphics[width=\textwidth]{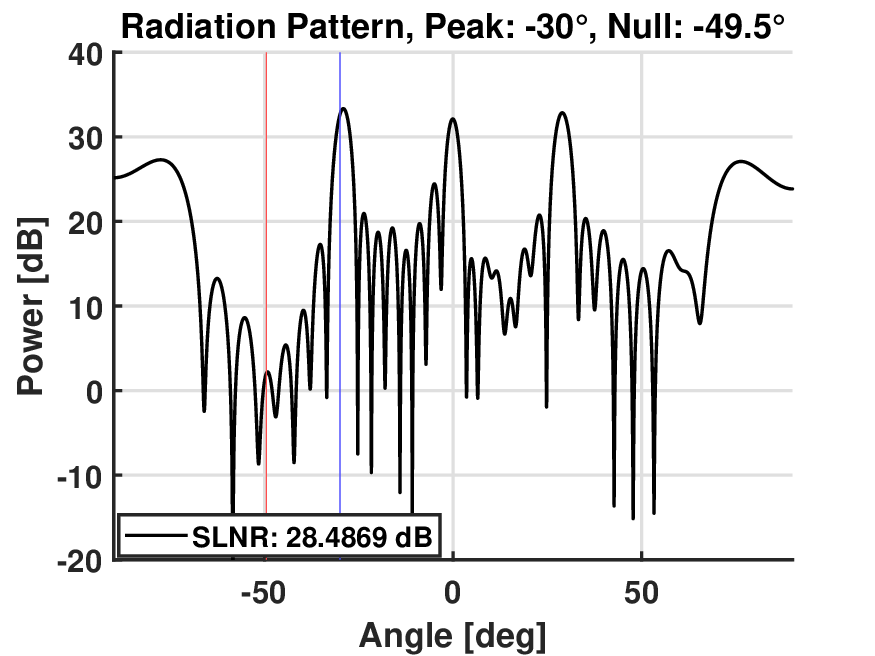}
        \caption{Beam at $-30\degree$, null at $-49.5\degree$. SLNR: 28.4869 dB.}
        \label{fig:rad_patterns_e}
    \end{subfigure}
    \hfill
    \begin{subfigure}[t]{0.24\textwidth}
        \includegraphics[width=\textwidth]{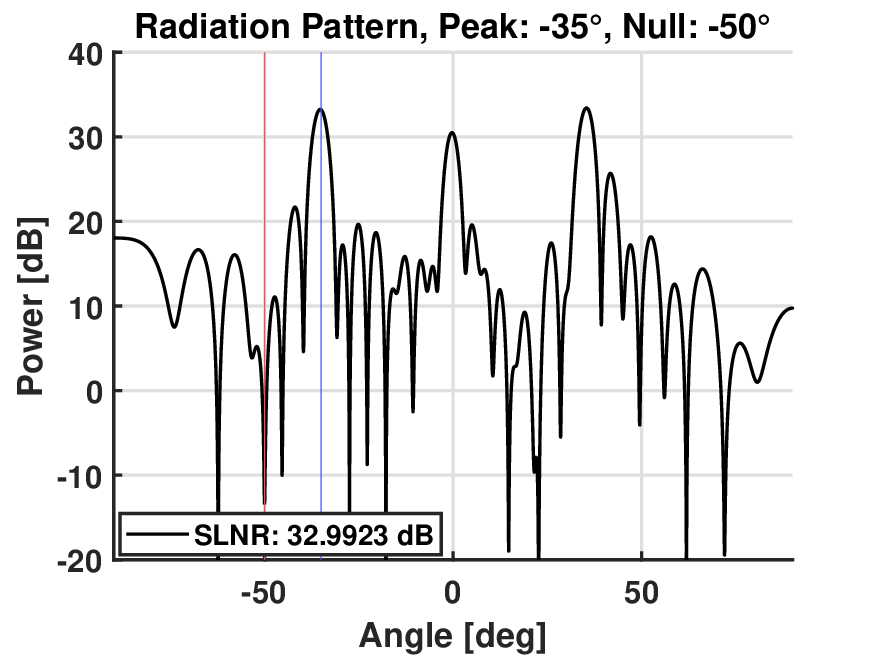}
        \caption{Beam at $-35\degree$, null at $-50\degree$. SLNR: 32.9923 dB.}
        \label{fig:rad_patterns_f}
    \end{subfigure}
    \hfill
    \begin{subfigure}[t]{0.24\textwidth}
        \includegraphics[width=\textwidth]{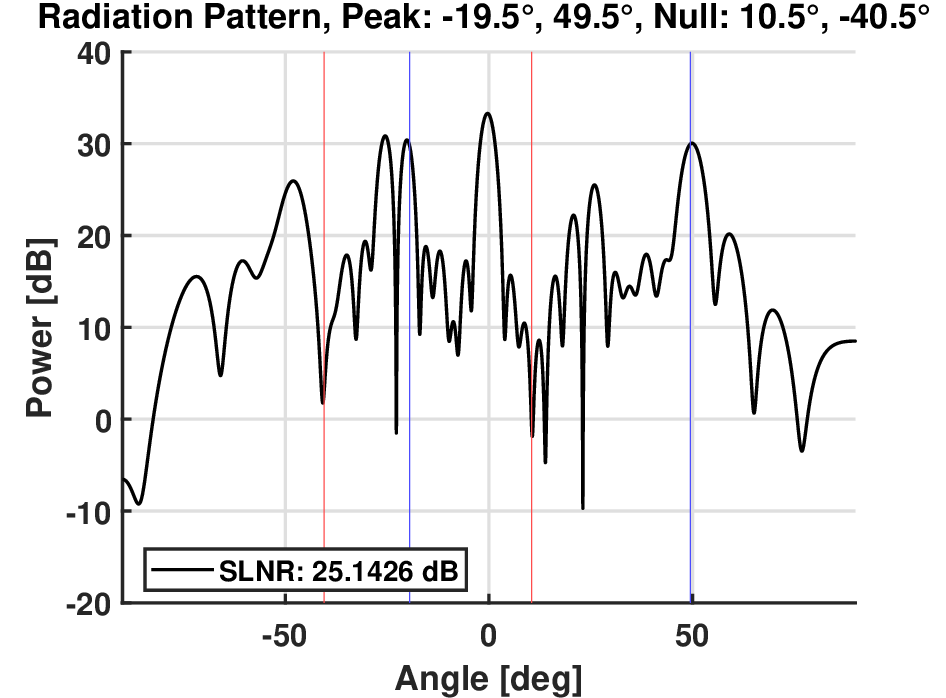}
        \caption{Beams at $-19.5\degree$ and $49.5\degree$, nulls at $10.5\degree$ and $-40.5\degree$. SLNR: 25.1426 dB.}
        \label{fig:rad_patterns_g}
    \end{subfigure}
    \hfill
    % \hspace{1cm}
    \begin{subfigure}[t]{0.24\textwidth}
        \includegraphics[width=\textwidth]{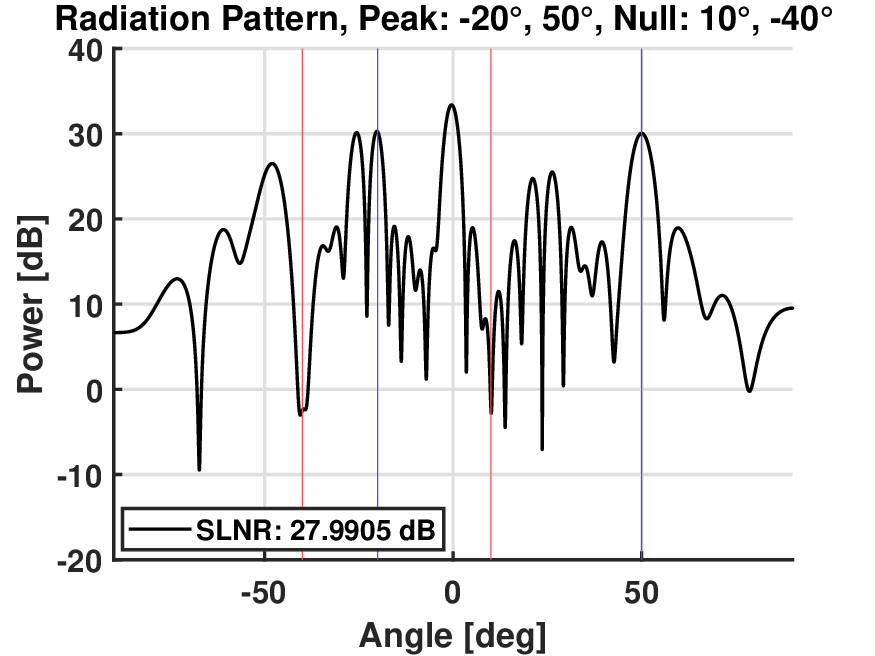}
        \caption{Beams at $-20\degree$ and $50\degree$, nulls at $10\degree$ and $-40\degree$. SLNR: 27.9905 dB.}
        \label{fig:rad_patterns_h}
    \end{subfigure}
    \hfill
\caption{Radiation patterns generated by the optimized NN, using SA and the lookup table.}
\label{fig:rad_patterns}
\end{figure*}

We analyze the performance of SA assisted by the lookup table in Table~\ref{table:lookup}. It is evident that the starting points created in Algorithm~\ref{alg:lookup} significantly improve the SLNR in nearly all cases, allowing the offline optimization using the NN to achieve superior performance over the online optimization in the simulated system. Furthermore, the standard deviations of the SLNR values are significantly smaller than before, indicating that the lookup table-based initializations reduce the risk that SA will not converge to a good solution. To further illustrate this result, all test cases are plotted in Fig.~\ref{fig:rad_patterns}.

\begin{algorithm}
    \caption{Adaptive Optimization and Lookup Table Construction}
    \label{alg:lookup}
    \begin{algorithmic}[1]
    \State $P_{\text{max}} \gets -\infty$
    \State $\boldsymbol{W} \gets [0,0,\ldots,0]^T$
    \If {lookup table does not contain any of $\boldsymbol{\theta}_{\text{beam}}$}
        \For {each $\theta_b$ in $\boldsymbol{\theta}_{\text{beam}}$}
            \State $P_b \gets $ max power~(\ref{eq:p_int}) of $\theta_b$ in training dataset.
            \If {$P_b > P_{\text{max}}$}
                \State $P_{\text{max}} \gets P_b$
                \State {$\boldsymbol{W}\gets \boldsymbol{W}_b$ (set of BSW amplitudes corresponding to $P_b$ from dataset).}
            \EndIf
        \EndFor
        \State {Define an empty objective function $\mathcal{C}$ based on the SLNR measure from~(\ref{eq:slnr}), initially not taking into account any beam or null directions.}
        \For {each $\theta_b$ in $\boldsymbol{\theta}_{\text{beam}}$}
            \State {Update $\mathcal{C}$ to include the SNR corresponding to beam direction $\theta_b$ in the numerator of~(\ref{eq:slnr}).}
            \State {Optimize $\boldsymbol{W}$ using SA to maximize $\mathcal{C}$.}
            \State {Store $\boldsymbol{W}$, beam directions from $\mathcal{C}$, and SLNR in the lookup table.}
        \EndFor
        \State {Update $\mathcal{C}$ to include the SNRs corresponding to all null directions $\boldsymbol{\theta}_{\text{null}}$, if any, in the denominator of~(\ref{eq:slnr}}).
        \State {Optimize $\boldsymbol{W}$ using SA to maximize $\mathcal{C}$.}
        \State {Store $\boldsymbol{W}$, beam and null directions from $\mathcal{C}$, and SLNR in the lookup table.}
    \Else
        \State {Find the $\boldsymbol{W}$ that creates peaks in the directions from the largest subset $\boldsymbol{\theta}_{\text{s},b}$ of $\boldsymbol{\theta}_{\text{beam}}$ from the lookup table. If two or more subsets with the same cardinality exist, choose the one with the highest SLNR and initialize $\mathcal{C}$ to the corresponding objective function with its beam and null directions.}
        \State {Perform the steps in lines 12-19, starting with $\mathcal{C}$ from line 21, iterating through all $\theta_b \in \{\boldsymbol{\theta}_{\text{beam}} \setminus \boldsymbol{\theta}_{\text{s},b}\}$ and then all $\boldsymbol{\theta}_{\text{null}}$, if any.}
    \EndIf
    \end{algorithmic}
\end{algorithm}

\begin{table*}
    \centering
    \begin{tabular}{|c|c|c|c|l||c|c|c|c|}
    \hline
    \multicolumn{5}{|c||}{} & 
    \multicolumn{4}{c|}{\textbf{NN System}}
    \\ \hline
     Beam 1 & Beam 2 & Null 1 & Null 2 & & Avg & Min & Max & Std. dev \\ \hline
     \multirow{2}*{-40.5} & \multirow{2}*{-} & \multirow{2}*{-} & \multirow{2}*{-} & Simulated & 33.6676 & 33.6292 & 33.7279 & 0.0361 \\ \cline{5-9}
     & & & & Estimated & 35.9271 & 35.8175 & 36.0131 & 0.0503 \\ \hline
     \multirow{2}*{25.5} & \multirow{2}*{-} & \multirow{2}*{-} & \multirow{2}*{-} & Simulated & 33.9228 & 33.8352 & 33.9872 & 0.0549 \\ \cline{5-9}
     & & & & Estimated & 36.9640 & 36.8946 & 37.0644 & 0.0615 \\ \hline
     \multirow{2}*{-30} & \multirow{2}*{-} & \multirow{2}*{-49.5} & \multirow{2}*{-} & Simulated & 27.6829 & 26.5275 & 28.4869 & 0.6647 \\ \cline{5-9}
     & & & & Estimated & 34.2191 & 33.9985 & 34.4114 & 0.1206 \\ \hline
     \multirow{2}*{-19.5} & \multirow{2}*{-49.5} & \multirow{2}*{-} & \multirow{2}*{-} & Simulated & 28.3460 & 27.2197 & 28.9696 & 0.7212 \\ \cline{5-9}
     & & & & Estimated & 28.6516 & 27.6357 & 29.2950 & 0.6165 \\ \hline
     \multirow{2}*{-19.5} & \multirow{2}*{49.5} & \multirow{2}*{10.5} & \multirow{2}*{-40.5} & Simulated & 23.0232 & 20.5317 & 25.1426 & 1.6340 \\ \cline{5-9}
     & & & & Estimated & 27.9592 & 27.6960 & 28.3134 & 0.2082 \\ \hline
     \multirow{2}*{30} & \multirow{2}*{-} & \multirow{2}*{0} & \multirow{2}*{-} & Simulated & 5.3311 & 4.7714 & 5.9562 & 0.3877 \\ \cline{5-9}
     & & & & Estimated & 7.5584 & 7.4377 & 7.6830 & 0.0720 \\ \hline
     \multirow{2}*{-35} & \multirow{2}*{-} & \multirow{2}*{-50} & \multirow{2}*{-} & Simulated & 31.4067 & 29.3688 & 32.9923 & 1.2463 \\ \cline{5-9}
     & & & & Estimated & 32.4866 & 32.3519 & 32.6637 & 0.1030 \\ \hline
     \multirow{2}*{-20} & \multirow{2}*{50} & \multirow{2}*{10} & \multirow{2}*{-40} & Simulated & 25.4036 & 21.8868 & 27.9905 & 1.7480 \\ \cline{5-9}
     & & & & Estimated & 26.1828 & 25.2871 & 26.7366 & 0.4128 \\ \hline
    \end{tabular}
    \caption{SLNR [dB] After Convergence of Simulated Annealing for the NN System Using the Lookup Table.}
    \label{table:lookup}
\end{table*}

\begin{figure}
    \centering
    \includegraphics[width=0.8\linewidth]{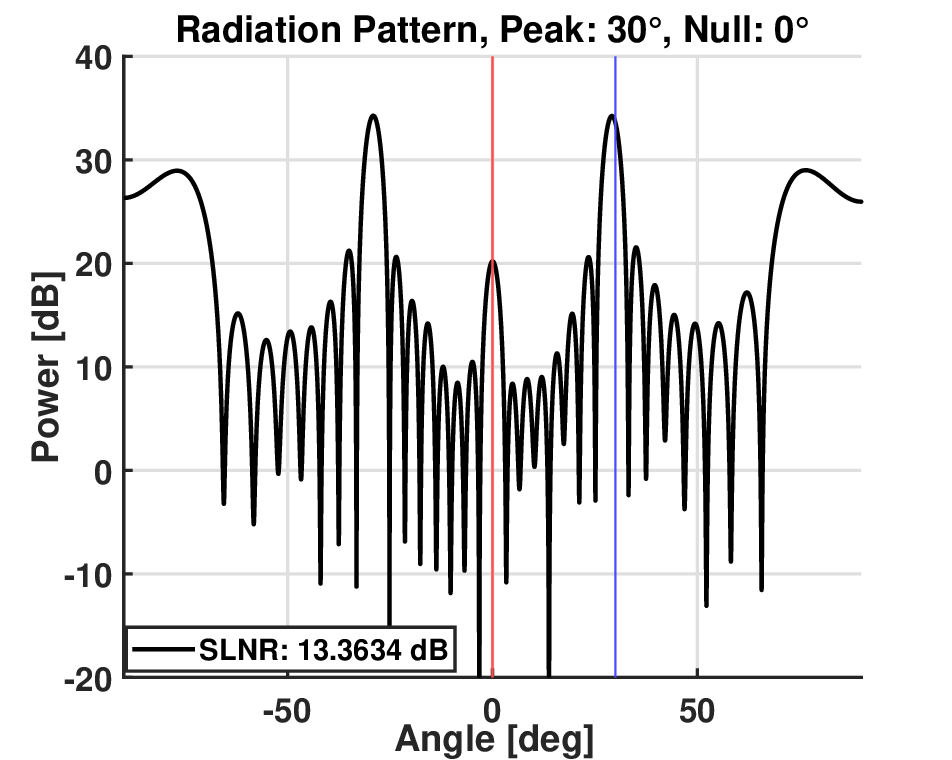}
    \caption{Generating a peak at $30\degree$ using SA and a dominant BSW at 3.5V. SNR at $30\degree$ is 33.5964 dB, peak at $0\degree$ is 13.3634 dB lower than the desired peak.}
    \label{fig:p30_higher_voltage}
\end{figure}

\begin{figure}
    \centering
    \includegraphics[width=0.8\linewidth]{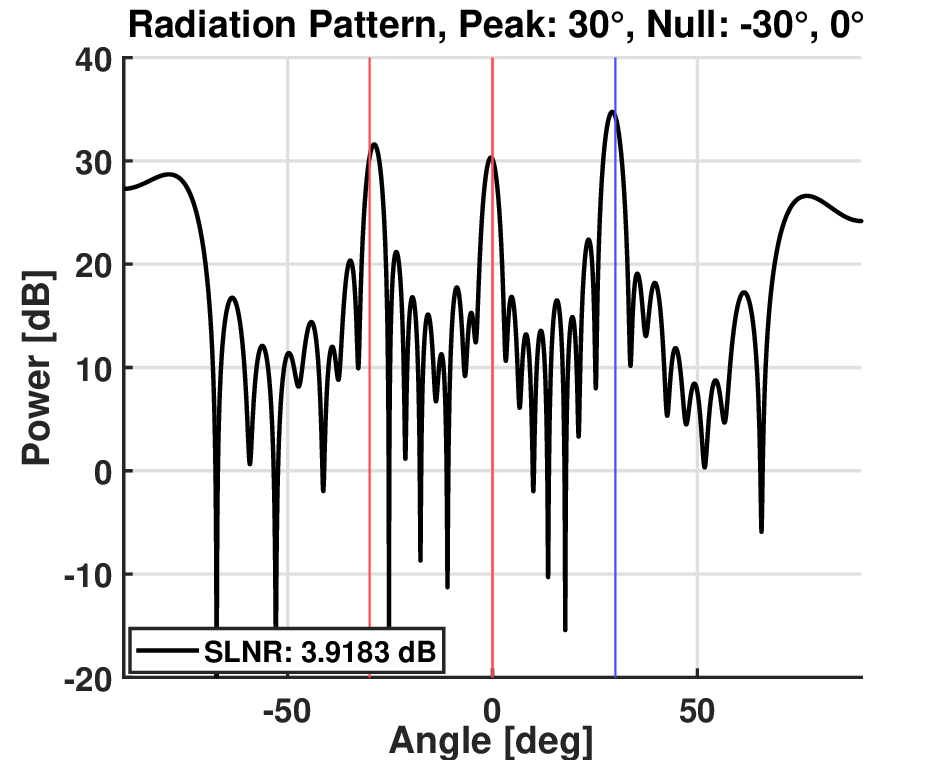}
    \caption{Generating a peak at $30\degree$ using SA and $N=50$ BSWs instead of $N=25$. SNR at $30\degree$ is 34.2803 dB, symmetric peak at $-30\degree$ is attenuated by 3.9183 dB.}
    \label{fig:p30_more_modes}
\end{figure}

It is observed from the plots that the offline and adaptive data-based optimizations successfully create radiation patterns that satisfy the criteria defined by the objective functions, creating peaks and nulls at the desired directions. However, peaks are also created in unintended directions, especially in the broadside ($0\degree$) direction and at angles symmetric with the desired beams, which may lead to undesirable behavior in a given environment \cite{10633724}. This is at least partially due to the use of the SLNR measure, which does not penalize radiation patterns with spurious beams. But a careful choice of the parameters used when creating the training dataset for the machine learning model can reduce such artifacts. For example, taking the case of Fig.~\ref{fig:rad_patterns_b} with a single desired beam at $30\degree$, it was found that setting the dominant BSW amplitude in an optimal $\boldsymbol{W}$ array to 3.5 V yields a higher peak at $30\degree$ with more attenuation at $0\degree$, as seen in Fig.~\ref{fig:p30_higher_voltage}. The ML model was likely unable to explore this scenario since this value is too close to the maximum value the model was trained on. Thus, there is a low probability of such a high value appearing frequently enough in the dataset for its effects to be accurately modeled. Moreover, by increasing the maximum number of BSW frequencies from $N=25$ to $N=50$, more degrees of freedom are available for beam pattern optimization at higher frequencies. This also helps reduce the reflected power of the symmetric beam at $-30\degree$, as shown in Fig.~\ref{fig:p30_more_modes}. These observations lead to the following conclusions:
\begin{itemize}
    \item \textbf{The voltage range of the randomly generated BSW amplitudes used for training should be carefully chosen depending on the desired performance of the RIS}. If the requirement from the ML model is to explore a wide range of radiation patterns, one should use relatively low random voltage amplitudes such as those used in this paper to avoid saturating the varactor voltage limits while allowing more BSWs to be activated at once. On the contrary, one can generate input arrays where fewer BSWs are excited, but with higher amplitudes, that will dominate the RIS biasing profile. This approach may also reduce the amount of data required for model training.%On the contrary, if the RIS is designed for simple cases like creating a single beam, higher amplitudes should be generated to isolate the relationships of fewer BSWs with their corresponding radiation patterns.
    \item \textbf{The number of BSWs and their frequencies should be chosen depending on the hardware constraints and the complexity of the system model.} Using more BSWs creates a tradeoff between the increased control over the radiation patterns generated by the RIS, and the higher-dimensional optimization problems, potentially increased amount of data collection required when generating an ML model, hardware complexity of the RIS, and control signal overhead.% This is a tradeoff when attempting to boost the overall RIS performance.
\end{itemize}

An inherent weakness of using a lookup table to store optimal RIS configurations is that its size increases exponentially with the number of beam pattern constraints. Possible solutions to this problem include interpolation between configurations in the lookup table, using for example manifold methods \cite{zimmermann2021manifold}, empirical interpolation \cite{doi:10.1137/090766498}, or additional ML models to reduce the memory requirements of the lookup table and the required number of SA executions for beam pattern optimization.

\section{CONCLUSION}
We have presented a novel data-driven approach for optimizing wave-controlled RIS without requiring extensive modeling of the RIS behavior or channel state information. The RIS is simulated in an environment where the channel models are unknown. It is excited with random combinations of BSW amplitudes, and received powers are sampled at various directions of interest for each set of input BSWs. An NN that generalizes the relationship between the BSW amplitudes and the corresponding radiation patterns is designed and optimized using a GA, achieving low training and validation losses on the collected data. SA is then used for offline beam pattern optimization using the NN-generated model. The optimal BSW weights are then used to excite the simulated RIS to create the desired radiation patterns. In conjunction, a lookup table is used to store optimal sets of BSW amplitudes to rapidly determine the biasing voltages for standard beam patterns and to create baselines for more complicated radiation patterns. The combination of the lookup table and SA was demonstrated to be capable of achieving superior performance in terms of SLNR in both sampled and interpolated directions via offline optimization using the NN, compared to simulated scenarios where live feedback from the receivers is used to determine convergence for SA.

The GA that was used in this paper to optimize the NN architecture for RIS modeling is applicable to many other optimization and modeling problems. The optimal NN architecture created for our application resembles an autoencoder that learns the general features from the BSW amplitude combinations in the first few hidden layers and maps them to the resulting radiation pattern as the information propagates through subsequent layers. Designing a good NN architecture to model high-dimensional nonlinear mappings is often not an intuitive or a straightforward task, requiring significant tuning and verification. This is especially crucial for regression problems where the model must generalize well for both the training and validation data. The GA was proven to be effective in automating the tuning process and determining system parameters that allow the NN to accurately represent the given model. 
%The SA algorithm successfully escapes local minima with minimal computational intensity, as it relies on addition and subtraction of randomized values to the given input vector, and may be the best algorithm for these purposes for many cases. 
Overall, the combination of backpropagation, GA, and SA was shown to lead to desirable RIS performance in the absence of an accurate physical model for the RIS or the environment, and provides a promising avenue for other RIS-related applications. 
\bibliographystyle{IEEEtran}
\bibliography{IEEEabrv,bibJournalList,bibfile}

% Generated by IEEEtran.bst, version: 1.14 (2015/08/26)
\begin{thebibliography}{10}
\providecommand{\url}[1]{#1}
\csname url@samestyle\endcsname
\providecommand{\newblock}{\relax}
\providecommand{\bibinfo}[2]{#2}
\providecommand{\BIBentrySTDinterwordspacing}{\spaceskip=0pt\relax}
\providecommand{\BIBentryALTinterwordstretchfactor}{4}
\providecommand{\BIBentryALTinterwordspacing}{\spaceskip=\fontdimen2\font plus
\BIBentryALTinterwordstretchfactor\fontdimen3\font minus
  \fontdimen4\font\relax}
\providecommand{\BIBforeignlanguage}[2]{{%
\expandafter\ifx\csname l@#1\endcsname\relax
\typeout{** WARNING: IEEEtran.bst: No hyphenation pattern has been}%
\typeout{** loaded for the language `#1'. Using the pattern for}%
\typeout{** the default language instead.}%
\else
\language=\csname l@#1\endcsname
\fi
#2}}
\providecommand{\BIBdecl}{\relax}
\BIBdecl

\bibitem{9475160}
C.~Pan, H.~Ren, K.~Wang, J.~F. Kolb, M.~Elkashlan, M.~Chen, M.~Di~Renzo,
  Y.~Hao, J.~Wang, A.~L. Swindlehurst, X.~You, and L.~Hanzo, ``{Reconfigurable
  Intelligent Surfaces for 6G Systems: Principles, Applications, and Research
  Directions},'' \emph{IEEE Communications Magazine}, vol.~59, no.~6, pp.
  14--20, 2021.

\bibitem{6331512}
F.~Capolino, A.~Vallecchi, and M.~Albani, ``{Equivalent Transmission Line Model
  With a Lumped X-Circuit for a Metalayer Made of Pairs of Planar
  Conductors},'' \emph{IEEE Transactions on Antennas and Propagation}, vol.~61,
  no.~2, pp. 852--861, 2013.

\bibitem{9887248}
D.~Hanna, M.~S. Melo, F.~Shan, and F.~Capolino, ``{A Versatile Polynomial Model
  for Reflection by a Reflective Intelligent Surface with Varactors},'' in
  \emph{Proc. IEEE International Symposium on Antennas and Propagation and
  USNC-URSI Radio Science Meeting (AP-S/URSI)}, 2022, pp. 679--680.

\bibitem{9770088}
E.~Ayanoglu, F.~Capolino, and A.~L. Swindlehurst, ``{Wave-Controlled
  Metasurface-Based Reconfigurable Intelligent Surfaces},'' \emph{IEEE Wireless
  Communications}, vol.~29, no.~4, pp. 86--92, 2022.

\bibitem{10742896}
G.~Ben-Itzhak, M.~Saavedra-Melo, B.~Bradshaw, E.~Ayanoglu, F.~Capolino, and
  A.~Lee~Swindlehurst, ``{Design and Operation Principles of a Wave-Controlled
  Reconfigurable Intelligent Surface},'' \emph{IEEE Open Journal of the
  Communications Society}, vol.~5, pp. 7730--7751, 2024.

\bibitem{10361836}
H.~Zhou, M.~Erol-Kantarci, Y.~Liu, and H.~V. Poor, ``{A Survey on Model-Based,
  Heuristic, and Machine Learning Optimization Approaches in RIS-Aided Wireless
  Networks},'' \emph{IEEE Communications Surveys \& Tutorials}, vol.~26, no.~2,
  pp. 781--823, 2024.

\bibitem{9380744}
X.~Hu, C.~Masouros, and K.-K. Wong, ``{Reconfigurable Intelligent Surface Aided
  Mobile Edge Computing: From Optimization-Based to Location-Only
  Learning-Based Solutions},'' \emph{IEEE Transactions on Communications},
  vol.~69, no.~6, pp. 3709--3725, 2021.

\bibitem{9448826}
Y.~Song, M.~R.~A. Khandaker, F.~Tariq, K.-K. Wong, and A.~Toding, ``{Truly
  Intelligent Reflecting Surface-Aided Secure Communication Using Deep
  Learning},'' in \emph{2021 IEEE 93rd Vehicular Technology Conference
  (VTC2021-Spring)}, 2021, pp. 1--6.

\bibitem{9317827}
B.~Sheen, J.~Yang, X.~Feng, and M.~M.~U. Chowdhury, ``{A Deep Learning Based
  Modeling of Reconfigurable Intelligent Surface Assisted Wireless
  Communications for Phase Shift Configuration},'' \emph{IEEE Open Journal of
  the Communications Society}, vol.~2, pp. 262--272, 2021.

\bibitem{8955968}
J.~Gao, C.~Zhong, X.~Chen, H.~Lin, and Z.~Zhang, ``{Unsupervised Learning for
  Passive Beamforming},'' \emph{IEEE Communications Letters}, vol.~24, no.~5,
  pp. 1052--1056, 2020.

\bibitem{10237459}
M.~Saavedra-Melo, K.~Rouhi, and F.~Capolino, ``{Wave-Controlled RIS: A Novel
  Method for Reconfigurable Elements Biasing},'' in \emph{Proc. IEEE
  International Symposium on Antennas and Propagation and USNC-URSI Radio
  Science Meeting (USNC-URSI)}, 2023, pp. 979--980.

\bibitem{Costa21}
F.~Costa and M.~Borgese, ``{Electromagnetic Model of Reflective Intelligent
  Surfaces},'' \emph{IEEE Open Journal of the Communications Society}, vol.~2,
  pp. 1577--1589, 2021.

\bibitem{Sievenpiper99}
D.~Sievenpiper, L.~Zhang, R.~Broas, N.~Alexopolous, and E.~Yablonovitch,
  ``{High-Impedance Electromagnetic Surfaces with a Forbidden Frequency
  Band},'' \emph{IEEE Trans. on Microwave Theory and Techniques}, vol.~47,
  no.~11, pp. 2059--2074, 1999.

\bibitem{Best08}
S.~R. Best and D.~L. Hanna, ``{Design of a Broadband Dipole in Close Proximity
  to an {EBG} Ground Plane},'' \emph{IEEE Antennas and Propagation Magazine},
  vol.~50, no.~6, pp. 52--64, 2008.

\bibitem{929650}
R.~Achar and M.~Nakhla, ``{Simulation of High-Speed Interconnects},''
  \emph{Proceedings of the IEEE}, vol.~89, no.~5, pp. 693--728, 2001.

\bibitem{sedra2020microelectronic}
A.~Sedra, K.~Smith, T.~Carusone, and V.~Gaudet, \emph{{Microelectronic
  Circuits}}, ser. Oxford Series in Electrical and Computer Engineering.\hskip
  1em plus 0.5em minus 0.4em\relax Oxford University Press, 2020.

\bibitem{9319694}
G.~Gradoni and M.~Di~Renzo, ``{End-to-End Mutual Coupling Aware Communication
  Model for Reconfigurable Intelligent Surfaces: An Electromagnetic-Compliant
  Approach Based on Mutual Impedances},'' \emph{IEEE Wireless Communications
  Letters}, vol.~10, no.~5, pp. 938--942, 2021.

\bibitem{7870611}
M.~Yazdi and M.~Albooyeh, ``{Analysis of Metasurfaces at Oblique Incidence},''
  \emph{IEEE Transactions on Antennas and Propagation}, vol.~65, no.~5, pp.
  2397--2404, 2017.

\bibitem{9530717}
S.~Basharat, S.~A. Hassan, H.~Pervaiz, A.~Mahmood, Z.~Ding, and M.~Gidlund,
  ``{Reconfigurable Intelligent Surfaces: Potentials, Applications, and
  Challenges for 6G Wireless Networks},'' \emph{IEEE Wireless Communications},
  vol.~28, no.~6, pp. 184--191, 2021.

\bibitem{9328501}
X.~Wei, D.~Shen, and L.~Dai, ``{Channel Estimation for RIS Assisted Wireless
  Communications—Part I: Fundamentals, Solutions, and Future
  Opportunities},'' \emph{IEEE Communications Letters}, vol.~25, no.~5, pp.
  1398--1402, 2021.

\bibitem{8683663}
D.~Mishra and H.~Johansson, ``{Channel Estimation and Low-complexity
  Beamforming Design for Passive Intelligent Surface Assisted MISO Wireless
  Energy Transfer},'' in \emph{Proc. IEEE International Conference on
  Acoustics, Speech and Signal Processing (ICASSP)}, 2019, pp. 4659--4663.

\bibitem{9771077}
A.~L. Swindlehurst, G.~Zhou, R.~Liu, C.~Pan, and M.~Li, ``{Channel Estimation
  With Reconfigurable Intelligent Surfaces -- A General Framework},''
  \emph{Proceedings of the IEEE}, vol. 110, no.~9, pp. 1312--1338, 2022.

\bibitem{10.1007/978-3-540-45179-2_53}
V.~Franc and V.~Hlav{\'a}{\v{c}}, ``{Greedy Algorithm for a Training Set
  Reduction in the Kernel Methods},'' in \emph{Computer Analysis of Images and
  Patterns}, N.~Petkov and M.~A. Westenberg, Eds.\hskip 1em plus 0.5em minus
  0.4em\relax Berlin, Heidelberg: Springer Berlin Heidelberg, 2003, pp.
  426--433.

\bibitem{7547360}
Y.~Sun, P.~Babu, and D.~P. Palomar, ``{Majorization-Minimization Algorithms in
  Signal Processing, Communications, and Machine Learning},'' \emph{IEEE
  Transactions on Signal Processing}, vol.~65, no.~3, pp. 794--816, 2017.

\bibitem{mlp}
M.-C. Popescu, V.~Balas, L.~Perescu-Popescu, and N.~Mastorakis, ``{Multilayer
  Perceptron and Neural Networks},'' \emph{WSEAS Transactions on Circuits and
  Systems}, vol.~8, 07 2009.

\bibitem{HORNIK1989359}
K.~Hornik, M.~Stinchcombe, and H.~White, ``{Multilayer Feedforward Networks are
  Universal Approximators},'' \emph{Neural Networks}, vol.~2, no.~5, pp.
  359--366, 1989.

\bibitem{7731699}
M.~A.~J. Idrissi, H.~Ramchoun, Y.~Ghanou, and M.~Ettaouil, ``{Genetic Algorithm
  for Neural Network Architecture Optimization},'' in \emph{Proc. 3rd
  International Conference on Logistics Operations Management (GOL)}, 2016.

\bibitem{2523}
H.~Ramchoun, Y.~Ghanou, M.~Ettaouil, and M.~A.~J. Idrissi, ``{Multilayer
  Perceptron: Architecture Optimization and Training},'' \emph{International
  Journal of Interactive Multimedia and Artificial Intelligence}, vol.~4,
  no.~1, pp. 26--30, 09/2016 2016.

\bibitem{taghvaee2021radiation}
H.~Taghvaee, A.~Jain, X.~Timoneda, C.~Liaskos, S.~Abadal, E.~Alarc{\'o}n, and
  A.~Cabellos-Aparicio, ``{Radiation Pattern Prediction for Metasurfaces: A
  Neural Network-Based Approach},'' \emph{Sensors}, vol.~21, no.~8, p. 2765,
  2021.

\bibitem{589532}
J.~Sola and J.~Sevilla, ``{Importance of Input Data Normalization for the
  Application of Neural Networks to Complex Industrial Problems},'' \emph{IEEE
  Transactions on Nuclear Science}, vol.~44, no.~3, pp. 1464--1468, 1997.

\bibitem{smith2018disciplinedapproachneuralnetwork}
\BIBentryALTinterwordspacing
L.~N. Smith, ``{A Disciplined Approach to Neural Network Hyper-Parameters: Part
  1 -- Learning Rate, Batch Size, Momentum, and Weight Decay},'' 2018.
  [Online]. Available: \url{https://arxiv.org/abs/1803.09820}
\BIBentrySTDinterwordspacing

\bibitem{forrest1996genetic}
S.~Forrest, ``{Genetic Algorithms},'' \emph{ACM Computing Surveys (CSUR)},
  vol.~28, no.~1, pp. 77--80, 1996.

\bibitem{DOMASHOVA2021263}
J.~V. Domashova, S.~S. Emtseva, V.~S. Fail, and A.~S. Gridin, ``{Selecting an
  Optimal Architecture of Neural Network Using Genetic Algorithm},''
  \emph{Procedia Computer Science}, vol. 190, pp. 263--273, 2021, 2020 Annual
  International Conference on Brain-Inspired Cognitive Architectures for
  Artificial Intelligence: Eleventh Annual Meeting of the BICA Society.

\bibitem{8407425}
B.~Ding, H.~Qian, and J.~Zhou, ``{Activation Functions and Their
  Characteristics in Deep Neural Networks},'' in \emph{Proc. Chinese Control
  And Decision Conference (CCDC)}, 2018, pp. 1836--1841.

\bibitem{tensorflow2015-whitepaper}
\BIBentryALTinterwordspacing
M.~Abadi \emph{et~al.}, ``{TensorFlow: Large-Scale Machine Learning on
  Heterogeneous Systems},'' 2015. [Online]. Available:
  \url{https://www.tensorflow.org/}
\BIBentrySTDinterwordspacing

\bibitem{Shi2019}
G.~Shi, J.~Zhang, H.~Li, and C.~Wang, ``{Enhance the Performance of Deep Neural
  Networks via L2 Regularization on the Input of Activations},'' \emph{Neural
  Processing Letters}, vol.~50, no.~1, pp. 57--75, Aug 2019.

\bibitem{kingma2017adammethodstochasticoptimization}
\BIBentryALTinterwordspacing
D.~P. Kingma and J.~Ba, ``{Adam: A Method for Stochastic Optimization},'' 2017.
  [Online]. Available: \url{https://arxiv.org/abs/1412.6980}
\BIBentrySTDinterwordspacing

\bibitem{traveling-salesman-thermo}
V.~Cerny, ``{Thermodynamical Approach to the Traveling Salesman Problem: {An}
  Efficient Simulation Algorithm},'' \emph{Journal of Optimization Theory and
  Applications}, vol.~45, pp. 41--51, January 1985.

\bibitem{10633724}
M.~Merluzzi and A.~Clemente, ``{Anomalous and Specular Reflections of
  Reconfigurable Intelligent Surfaces: Configuration Strategies and System
  Performance},'' \emph{IEEE Wireless Communications Letters}, vol.~13, no.~10,
  pp. 2707--2711, 2024.

\bibitem{zimmermann2021manifold}
R.~Zimmermann, ``{Manifold Interpolation},'' \emph{Model Order Reduction},
  vol.~1, pp. 229--274, 2021.

\bibitem{doi:10.1137/090766498}
S.~Chaturantabut and D.~C. Sorensen, ``{Nonlinear Model Reduction via Discrete
  Empirical Interpolation},'' \emph{SIAM Journal on Scientific Computing},
  vol.~32, no.~5, pp. 2737--2764, 2010.

\end{thebibliography}
\vfill
\end{document}